# Early Release Science of the exoplanet WASP-39b with JWST NIRISS

Adina D. Feinstein[1,2,*], Michael Radica[3], Luis Welbanks[4,5], Catriona Anne Murray[6], Kazumasa Ohno[7], Louis-Philippe Coulombe[3], Néstor Espinoza[8,9], Jacob L. Bean[1], Johanna K. Teske[10], Björn Benneke[3], Michael R. Line[4], Zafar Rustamkulov[11], Arianna Saba[12], Angelos Tsiaras[12,13], Joanna K. Barstow[14], Jonathan J. Fortney[7], Peter Gao[10], Heather A. Knutson[15], Ryan J. MacDonald[5,16,17], Thomas Mikal-Evans[18], Benjamin V. Rackham[19,20,21], Jake Taylor[3,22], Vivien Parmentier[22,23], Natalie M. Batalha[7,24], Zachory K. Berta-Thompson[6], Aarynn L. Carter[7], Quentin Changeat[8,12,25], Leonardo A. Dos Santos[8], Neale P. Gibson[26], Jayesh M Goyal[27], Laura Kreidberg[18], Mercedes López-Morales[28], Joshua D Lothringer[29], Yamila Miguel[30,31], Karan Molaverdikhani[18,32,33], Sarah E. Moran[34], Giuseppe Morello[35,36,37], Sagnick Mukherjee[7], David K. Sing[9,11], Kevin B. Stevenson[38], Hannah R. Wakeford[39], Eva-Maria Ahrer[40,41], Munazza K. Alam[10], Lili Alderson[39], Natalie H. Allen[9,2], Natasha E. Batalha[42], Taylor J. Bell[43], Jasmina Blecic[44,45], Jonathan Brande[46], Claudio Caceres[47,48,49], S.L.Casewell[50], Katy L. Chubb[51], Ian J.M. Crossfield[46], Nicolas Crouzet[30], Patricio E. Cubillos[52,53], Leen Decin[54], Jean-Michel Désert[55], Joseph Harrington[56], Kevin Heng[32,41,57], Thomas Henning[18], Nicolas Iro[58], Eliza M.-R. Kempton[59], Sarah Kendrew[25], James Kirk[28,60,61], Jessica Krick[62], Pierre-Olivier Lagage[63], Monika Lendl[64], Luigi Mancini[18,65,66], Megan Mansfield[67,5], E. M. May[38], N. J. Mayne[68], Nikolay K. Nikolov[8], Enric Palle[35], Dominique J.M. Petit dit de la Roche[64], Caroline Piaulet[3], Diana Powell[28,5], Seth Redfield[69], Laura K. Rogers[70], Michael T. Roman[50,71], Pierre-Alexis Roy[3], Matthew C. Nixon[59,70], Everett Schlawin[67], Xianyu Tan[22], P. Tremblin[72], Jake D. Turner[17,5], Olivia Venot[73], William C. Waalkes[74,2], Peter J. Wheatley[40,41], Xi Zhang[75]

*Corresponding author's email: afeinstein@uchicago.edu
All author affiliations are listed at the end of the paper.

**Transmission spectroscopy provides insight into the atmospheric properties and consequently the formation history, physics, and chemistry of transiting exoplanets[1]. However, obtaining precise inferences of atmospheric properties from transmission spectra requires simultaneously measuring the strength and shape of multiple spectral absorption features from a wide range of chemical species[2–4]. This has been challenging given the precision and wavelength coverage of previous observatories[5]. Here, we present the transmission spectrum of the Saturn-mass exoplanet WASP-39 b obtained using the SOSS mode of the NIRISS instrument on the JWST. This spectrum spans 0.6–2.8 μm in wavelength and reveals multiple water absorption bands, the potassium resonance doublet, as well as signatures of clouds. The precision and broad wavelength coverage of NIRISS-SOSS allows us to break model degeneracies between cloud properties and the atmospheric composition of WASP-39 b, favouring a heavy element enhancement ("metallicity") of ~10 – 30× the solar value, a sub-solar carbon-to-oxygen (C/O) ratio, and**

**a solar-to-super-solar potassium-to-oxygen (K/O) ratio. The observations are also best explained by wavelength-dependent, non-gray clouds with inhomogeneous coverage of the planet's terminator.**

We observed a transit of WASP-39 b using the Near Infrared Imager and Slitless Spectrograph (NIRISS)[6] on the JWST as part of the Transiting Exoplanet Community Early Release Science Program (ERS)[7,8]. Our observations spanned 8.2 hours starting on UTC July 26, 2022 20:45, covering the 2.8-hour transit as well as 3.0 hours prior to and 2.4 hours after the transit to establish a flux baseline. The data were taken in the Single Object Slitless Spectroscopy (SOSS) mode, which simultaneously covers the wavelength range from $0.6 - 2.8$ µm across two spectral orders on the same detector. Order 1 contains the spectral range between $0.85 - 2.8$ µm at an average resolving power of $R \equiv \lambda/\Delta\lambda = 700$, while Order 2 delivers the spectral range of $0.6 - 1.4$ µm at an average resolving power of $R = 1400$. In the SOSS mode, the spectra are spread across more than 20 pixels in the cross-dispersion direction via a cylindrical defocusing lens (see Extended Data Fig. 1), thus allowing longer integration times and reducing the impact of pixel-level differences in the detector response. However, this defocus also results in the physical overlap of both orders on the detector, as well as the characteristic "horned" structure of the SOSS PSF. The time series observation was composed of 537 integrations of 49.4 seconds (nine groups per integration), corresponding to a duty cycle of 89%.

We extracted the stellar spectra from the time series observations using six different pipelines to test the impact of differences in spectral order tracing, $1/f$ noise correction, background removal, and spectrum extraction methodology (see Methods and Extended Data Figs. 2 and 3). For each pipeline, we created spectrophotometric light curves at either pixel resolution (that is one light curve per detector column) or the native instrument resolution (two to three pixels per bin) (Fig. 1). We also summed the data to create individual white-light curves for each order (Extended Data Fig. 4). The spectrophotometric and white-light curves are largely free of instrumental systematics except for a constant-rate linear trend in time and an exponential ramp effect within the first 15 minutes of the time series, the latter of which we trimmed from the remainder of the data analysis. The signal-to-noise in the spectrophotometric light curves for Order 1 at $1.34$µm is 165; the signal-to-noise for Order 2 at $0.71$ µm is 103.

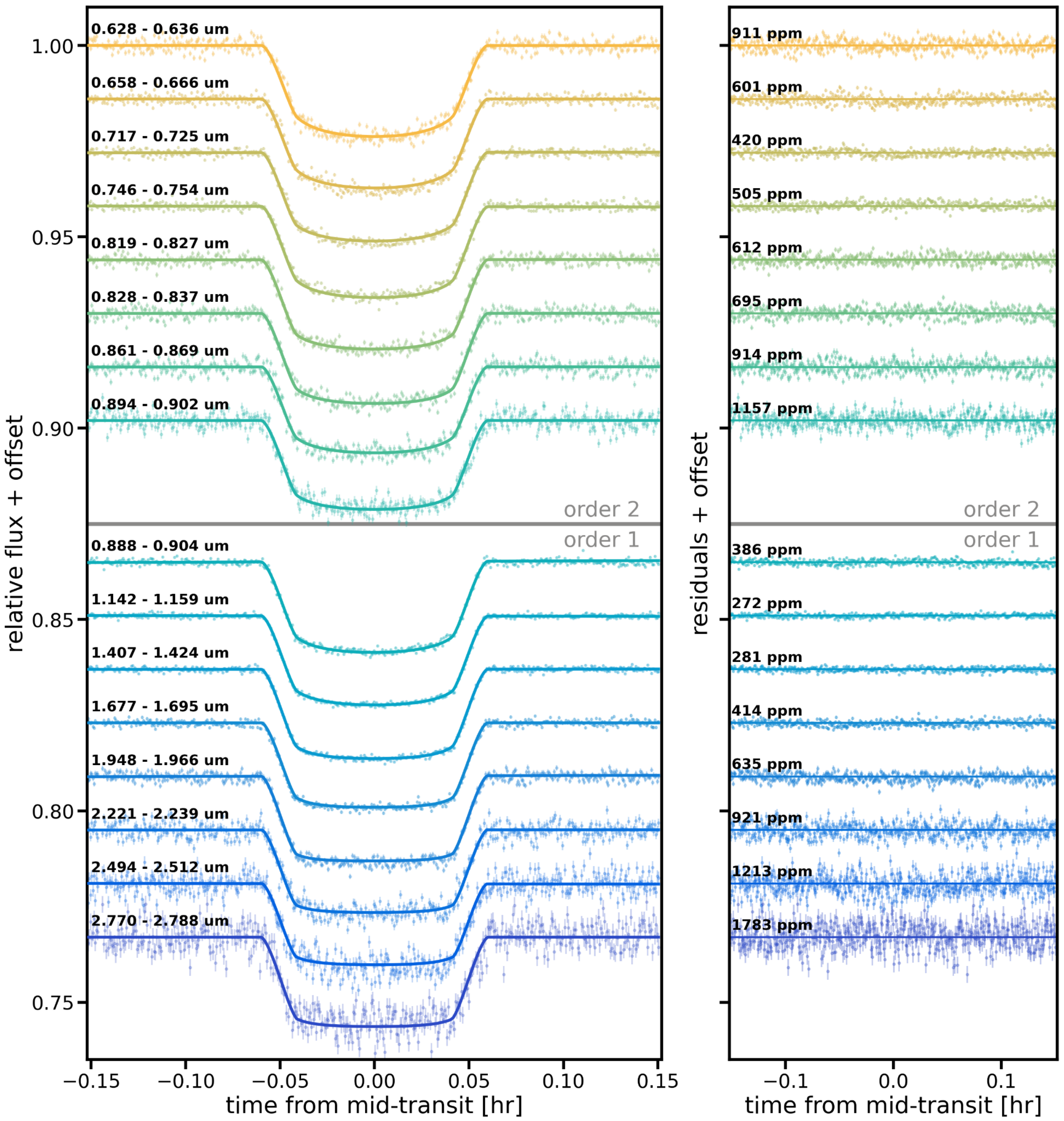

**Figure 1: Selection of spectrophotometric light curves and residuals for WASP-39b's transit observed with NIRISS-SOSS for Orders 1 and 2.** An exoplanet transit model (solid line) was fitted to each light curve with `chromatic_fitting` using a quadratic limb-darkening law. The limb-darkening coefficients, planet-to-star radius ratio ($R_p/R_*$), and out-of-transit flux were varied in each wavelength channel, while all other parameters were fixed. The residuals to the best-fit models are shown for each light curve. The wavelength range per each channel is denoted on the left plot, while parts-per-million (ppm) scatter in the residuals is denoted on the right plot. We calculate the ppm as the standard deviation of the out-of-transit residuals. The reductions are from the `nirHiss` and `chromatic_fitting` routines described in the Methods. (https://github.com/afeinstein20/wasp39b_niriss_paper/blob/main/scripts/figure1.py)

For each reduction (described in Methods), we fit the corresponding white-light curve per each order with a transit model to extract the best-fit orbital parameters (Extended Data Table 2). We fixed the orbital period $P$ to the best-fitting value from[9]. For the transit fits to the spectrophotometric light curves, most fitting routines also set the time of mid-transit $t_0$, impact parameter $b$, and semi-major axis to stellar radius ratio $a/R_*$ to the white-light parameters (Extended Data Table 2) to ensure consistency between reductions. We thus allowed only the planet-to-star radius ratio, $R_p/R_*$, and the two parameters of the quadratic limb-darkening law, $u_1$ and $u_2$, to vary for each wavelength, along with any additional systematics and long-term trends in the out-of-transit flux. These fits were carried out at either pixel resolution or the native instrument resolution depending on the pipeline. The high-resolution transit depths were subsequently binned into 100 spectral wavelength channels, 80 in Order 1 and 20 in Order 2, to create a transmission spectrum at $R \sim 300$. In Figure 2 we present this transmission spectrum of WASP-39b between $0.6 - 2.8\,\mu\mathrm{m}$ from the `nirHiss`, `supreme-SPOON`, and `transitspectroscopy` reduction pipelines. We find consistent results between the different data reduction pipelines and light curve fitting analyses, with the derived spectra also in agreement with previous results from the Hubble Space Telescope (HST, see also Extended Data Fig. 5).

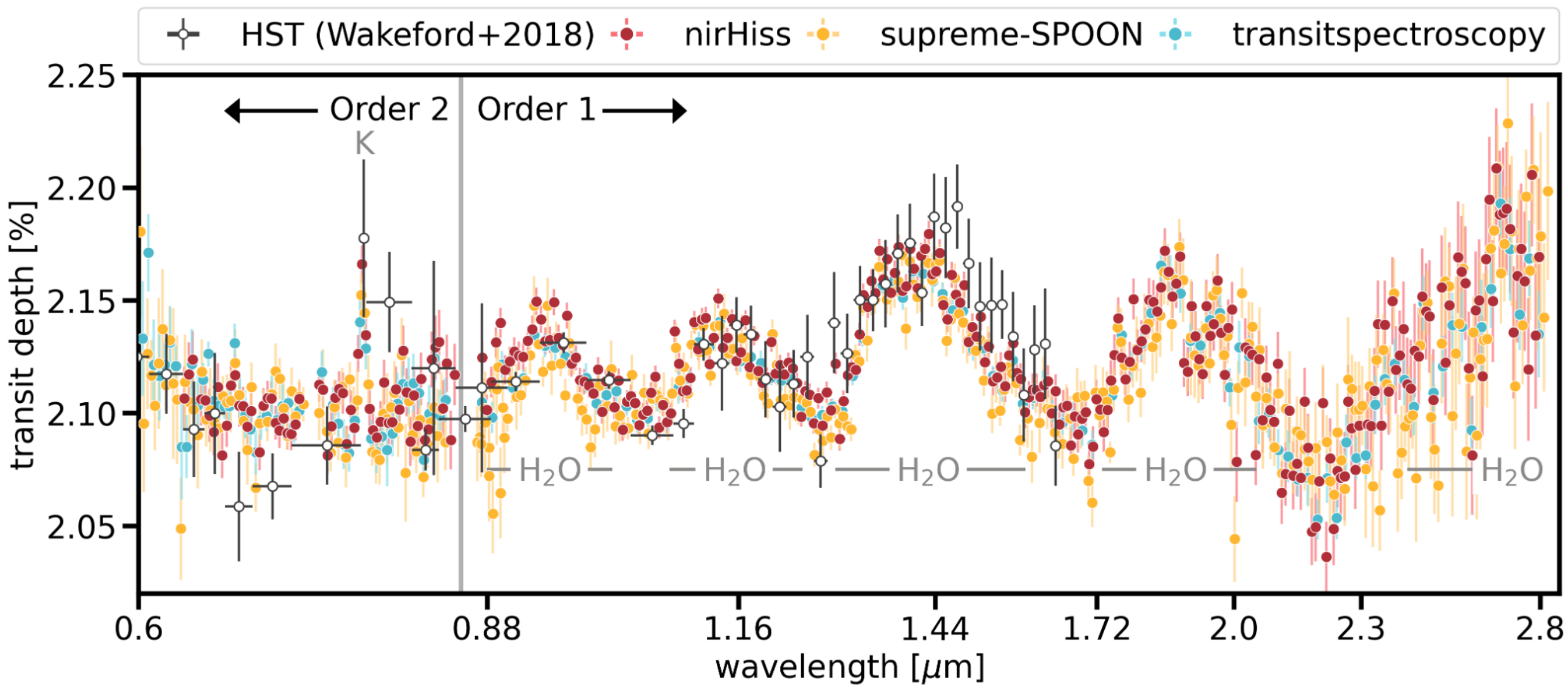

**Figure 2: NIRISS transmission spectra for WASP-39b obtained by three data reduction pipelines.** We find broad agreement in the overall structure of the transmission spectra between several reduction pipelines, a sample of which are presented here (see Extended Data Fig. 5 for all reductions). The JWST data are shown in the coloured points, while previous HST observations of WASP-39b[18] are shown in white. We note that we only consider wavelengths <0.85μm for Order 2, since Order 1 has much higher fidelity in the overlapping 0.85–1.0μm range. The JWST and HST data agree across the three broad $H_2O$ features that they have in common. We find evidence of a K absorption feature at 0.76μm, which the new JWST data, which was hypothesised in the previous HST data[18].
(https://github.com/afeinstein20/wasp39b_niriss_paper/blob/main/scripts/figure2.py)

We investigated the atmospheric properties of WASP-39b by comparing our measured transmission spectrum from the `nirHiss` pipeline to grids of one-dimensional, radiative-convective-thermochemical equilibrium models. These models explore the impact of atmospheric metallicity (M/H), carbon-to-oxygen ratio (C/O), potassium-to-oxygen ratio (K/O), heat redistribution (f), and cloud coverage on the transmission spectrum of the planet. We explored multiple cloud models ranging from parametric treatments[10,11] to a droplet sedimentation model[12] that calculates the vertical distributions of cloud mass mixing ratio and mean particle size from the balance between gravitational sedimentation and eddy diffusion of cloud particles. Using a Bayesian inference framework (see Methods), we compared these grids of models to the observations and inferred the range of M/H, C/O, K/O and f, that best explains the data while marginalising over different cloud treatments. WASP-39, the host star, has a metallicity equal to that of the Sun within measurement precision[13–16], so we reference the planet's atmospheric abundances to the solar pattern of elemental abundances[17]. We compared the grid spectra computed by various models (`PICASO`, `ATMO`, `Phoenix`, and `ScCHIMERA`) with an observational spectrum obtained from each data reduction pipeline and obtained broadly consistent results on the inferred atmospheric properties. In what follows, we introduce the results from the comparison between the spectrum from `nirHiss` data reduction and `ScCHIMERA` grid that allows the most comprehensive treatments of cloud properties.

Our best-fitting model to the NIRISS-SOSS transmission spectrum of WASP-39 b is presented in Figure 3. This model comes from the `ScCHIMERA` grid (see Methods) and assumes an atmospheric metallicity of [M/H]=1.38 ($23\times$ the solar value), C/O=0.2 ($0.55\times$ solar value,[17]), [K/O]=0.1 (1.26× the solar value), full day-night heat redistribution (f=1), and flux-balanced clouds with inhomogeneous terminator coverage. We label the unambiguous signatures of absorption due to potassium and water vapour in Figure 2. The spectral maxima at 0.9, 1.15, 1.4, and 1.8 μm due to water absorption result in a $> 30\sigma$ detection of the molecule (see Methods). Similarly, the potassium doublet at 0.768 μm is detected in the data at $6.8\sigma$. Besides these chemical species, signatures of CO and/or $CO_2$ are identified due to their contribution to the spectrum past $2.3$ μm, with a $3.6\sigma$ significance model preference for CO over $CO_2$ absorption (see Methods).

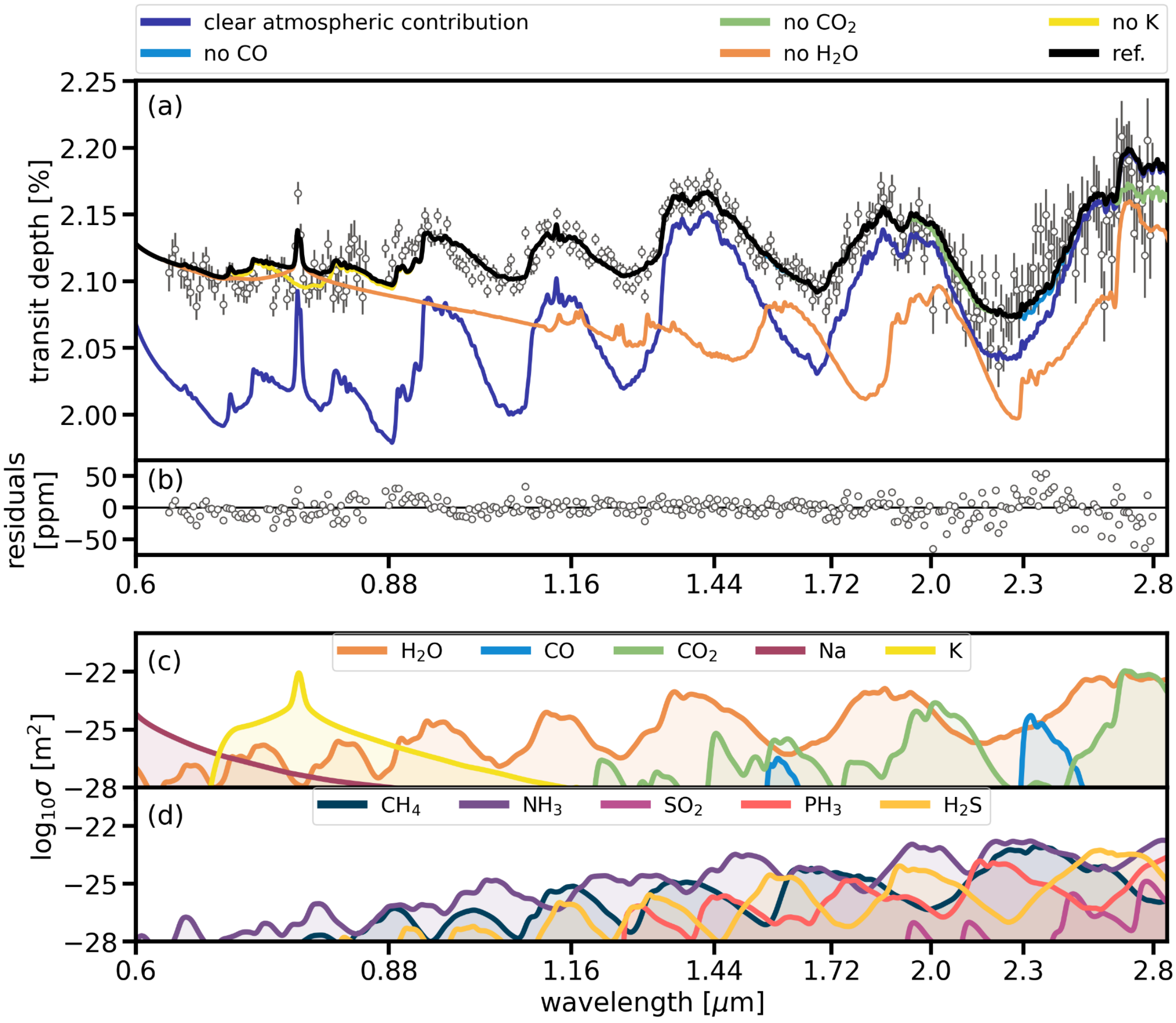

**Figure 3: Interpretation of the constituents of the NIRISS WASP-39b transmission spectrum. (a/b):** The top panel shows the comparison of WASP-39b's transmission spectrum from the `nirHiss` reduction (grey points) with respect to the best-fit reference model (black line). Each colored line removes a key constituent found in our best-fit reference model to demonstrate how the spectrum would change were these features not included. The removal of clouds, and $H_2O$ absorption from the reference model result in large-scale changes to the shape and depth of the transmission spectrum. Other sources of opacity with an impact on the spectrum are K, CO, and $CO_2$. Residuals between the data and the reference model are plotted below. **(c/d):** The two bottom panels show the molecular absorption cross-sections for a selection of gases observable within the NIRISS bandpass. The middle panel highlights gases inferred by our analysis of WASP-39b's spectrum. The bottom panel highlights some gases that were not identified in these data, but that may be present in future observations of other exoplanets. (https://github.com/afeinstein20/wasp39b_niriss_paper/blob/main/scripts/figure3.py)

From the chemical equilibrium models considered, we find that the observations are best explained by a sub-solar C/O (see Fig. 4, panel a). Across the different spectroscopic resolutions and atmospheric models, the best-fit C/O is 0.2, which is the lowest ratio explored in the grid of models. We rule out super-solar C/O due to the lack of $CH_4$ features at $\sim 1.7$ µm and $\sim 2.3$ µm, where they would be expected for C/O $\gtrsim 0.7$. Overall, solar-to-super-solar C/O ratios fail to explain the transmission spectrum at the shortest ($\lesssim$1 µm) and longest ($\gtrsim$2 µm) wavelengths, highlighting the importance of the continuous wavelength range in these NIRISS-SOSS observations. Our best-fit C/O is broadly consistent with the observations of WASP-39b with NIRCam ($2.4 - 4$µm; Ahrer et al. submitted), NIRSpec G395H ($3 - 5$µm; Alderson et al. submitted), and NIRSpec PRISM ($0.5 - 5$µm; Rustamkulov et al. submitted).

We also find that the observations are best explained by an atmospheric metallicity of $10–30\ \times$ solar. Metallicity inferences over the wavelength range of these observations are largely driven by the size and shape of the water vapour features, with some minor contributions due to CO and/or $CO_2$ at longer wavelengths ($>$ 2 µm; see Fig. 3 and Fig. 4, panel b). The preferred range of metallicities provides the best fit to the shape and size of the muted water vapour features shortward of 2 µm in combination with the larger water and CO/$CO_2$ feature longward of 2 µm, regardless of the assumed cloud treatment in our models.

Due to the simultaneous detection of potassium and water vapour, we are able to place constraints on the K/O ratio, which is a refractory-to-volatile elemental ratio, being a solar-to-super-solar value. Since the refractory elements are condensed into solids in most parts of protoplanetary disks, the disk gas accretion tends to cause a sub-stellar refractory elemental abundance[19]. By contrast, solid accretion, such as planetesimal accretion, acts to increase the refractory elemental abundance and refractory-to-volatile elemental ratio[20], though the latter depends on the composition of the accreted solids[21]. We anticipate that the K/O ratio diagnoses to what degree the solid accretion enriched the atmosphere during the formation stage. All of our fitted models find that the WASP-39b observations are well described by solar-to-super-solar K/O ratios, which is in agreement with previous inferences for this planet obtained via observations with limited spectral coverage[22]. We do not expect the K feature to be impacted by stellar chromospheric magnetic activity given the effective temperature of the star$\sim$ 5300K[23] and the general quietness of WASP-39 (see Ahrer et al. submitted) It is also in line with larger population studies of hot giant planets that broadly found solar-to-super-solar refractory abundances and solar-to-sub-solar $H_2O$ abundances[22,24]. The shape and strength of the potassium doublet are best explained by [K/O]~0.1–0.5, equivalent to 1–3× solar (see Extended Data Fig. 8). While the suggested K/O ratio might be a lower limit owing to the photoionization of K at upper atmospheres[25], these observations demonstrate the power of NIRISS-SOSS to obtain simultaneous inferences on novel atmospheric tracers beyond the well-established C/O ratio.

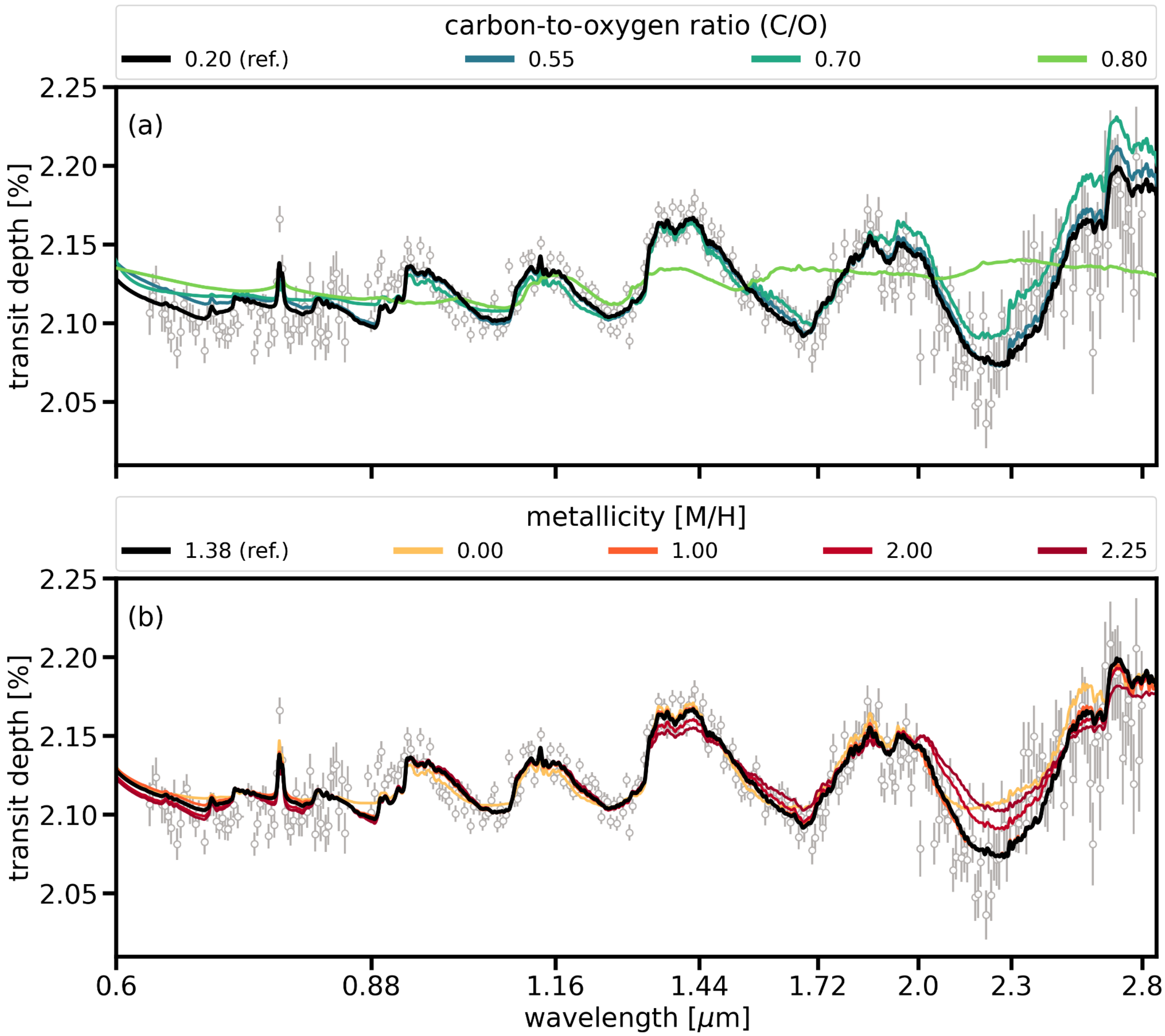

**Figure 4: Impact of the carbon-to-oxygen ratio (C/O) and metallicity on the JWST-NIRISS spectrum of WASP-39b. Top:** Variation of the C/O in the best-fit reference model, while keeping the metallicity, redistribution, and K/O parameters from the reference model the same, and fitting for the cloud parameters and scaled planetary radius to best explain the observations. Under these equilibrium conditions, increasing the C/O results in less $H_2O$ and more $CH_4$, the latter having spectroscopic signatures incompatible with the observations. To mute these incompatible $CH_4$ features at high C/O, the model requires a higher degree of cloudiness that also mutes any remaining $H_2O$ features in the spectrum. **Bottom:** The same exercise as above, but instead we vary the metallicity parameter. The metallicity constraint is driven by the $\lambda > 2\mu m$ data; the high-metallicity models ([M/H]> 2) expect larger transit depths than what is seen in the data. The same reference model is plotted as a thick black line in both panels.
(https://github.com/afeinstein20/wasp39b_niriss_paper/blob/main/scripts/figure4.py)

The NIRISS-SOSS observations enable the detection of clouds in the atmosphere of WASP-39b. Clear atmosphere models cannot explain the amplitudes of all of the water vapour features simultaneously, which strongly indicates the presence of clouds (see Methods and Extended Data Fig. 6). The atmospheric models explored here indicate the presence of non-gray and non-homogeneous clouds, with model preferences of $8\sigma$ and greater for models with both non-gray and non-homogeneous clouds over models with gray homogeneous clouds only. This model preference is driven by the decrease in transit depth between 2–2.3 μm (see Extended Data Fig. 7, panel a), which cannot be explained by gray clouds uniformly distributed along the terminator (see Extended Data Fig. 7, panel a). Moreover, within the various cloud treatments tested here (gray, gray + power-law, flux-balanced clouds, see Methods), both parametric and droplet sedimentation models indicate a preference for inhomogeneous cloud coverage of $\sim 50 - 70\%$ around the planetary day-night terminator because it better explains the decrease in transit depth between 2–2.3 μm. These results confirm the necessity for broad wavelength coverage, and information regarding the cloud coverage of the planet, to constrain the atmospheric metallicity of a planet[2,4,11]

Atmospheric circulation and cloud microphysical models have predicted that the cloud structure varies significantly along the terminators of hot Jupiters[26–28]. In particular, different compositions of clouds have different condensation temperatures and thus likely have different cloud coverage at the terminator[26]. Further studies combining temperature difference of east/west terminators to microphysical cloud models may be able to use the measured cloud coverage to determine the cloud composition of WASP-39b. Previous indications of non-gray or non-homogeneous clouds[5,29–33] have relied on a single or small number of spectroscopic points, making our inference here for WASP-39b of non-gray cloud with inhomogeneous terminator coverage in the transmission spectrum of an exoplanetary atmosphere the most confident to date. These constraints on the physical properties of clouds, alongside the multiple spectral features across a broad wavelength coverage, are key to breaking well-known degeneracies between the metallicity and cloud-cover in atmospheric models[2,4,11] and deriving constraints on the bulk atmospheric properties.

The precision and wavelength coverage of NIRISS-SOSS should allow us to obtain more precise and robust constraints on atmospheric composition and tracers of planet formation than most previous transmission spectroscopy observations. The super-solar metallicity of WASP-39 b and the solar-to-super-solar K/O are in agreement with previous studies of mass-metallicity trends in transiting exoplanets[18,22,34,35]. If confirmed with further detailed modelling, a super-solar K/O ratio in WASP-39 b's atmosphere would likely indicate enrichment due to the accretion of planetesimals[20–22], although the measurements of potassium and oxygen abundances for the host star are also needed in order to establish this result. Similarly, the suggestion of sub-solar C/O and super-solar metallicity may be compatible with a planetesimal accretion scenario[e.g., 36,37,38]. The combination of a super-solar metallicity, super-solar K/O ratio, and sub-solar C/O ratio may suggest the planet formed beyond the $H_2O$ snowline followed by inward migration, for which theory predicts efficient accretion of planetesimals at ~ 2–10 AU[e.g., 20,39]. At those orbital distances, the planetesimals likely contain K rock and $H_2O$ ice but almost no CO ice[e.g., 40,41], which explains the sub-solar C/O and super-solar K/O ratios along with a super-solar metallicity if a sufficient amount of planetesimals were accreted. However, fully understanding the possible formation pathways of this planet requires statistical constraints on the complete chemical inventory of the planet and the relative abundances of the carbon- oxygen- and alkali- bearing species. Such efforts will be possible when applying retrieval techniques to the complete transmission spectrum of WASP-39b from 0.5 to 5.5μm that is being produced by the Transiting Exoplanet Community ERS Program. Our results validate JWST's NIRISS-SOSS as an instrument mode capable of producing exquisite exoplanet atmosphere measurements and major advancements in our understanding of planet formation and exoplanet atmospheres.

1. Madhusudhan, N. Exoplanetary Atmospheres: Key Insights, Challenges, and Prospects. **57**, 617–663 (2019).
2. Benneke, B. & Seager, S. Atmospheric Retrieval for Super-Earths: Uniquely Constraining the Atmospheric Composition with Transmission Spectroscopy. **753**, 100 (2012).
3. Kempton, E. M.-R., Lupu, R., Owusu-Asare, A., Slough, P. & Cale, B. Exo-Transmit: An Open-Source Code for Calculating Transmission Spectra for Exoplanet Atmospheres of Varied Composition. **129**, 044402 (2017).
4. Welbanks, L. & Madhusudhan, N. On Degeneracies in Retrievals of Exoplanetary Transmission Spectra. **157**, 206 (2019).
5. Sing, D. K. *et al.* A continuum from clear to cloudy hot-Jupiter exoplanets without primordial water depletion. **529**, 59–62 (2016).
6. Doyon, R. *et al.* The JWST Fine Guidance Sensor (FGS) and Near-Infrared Imager and Slitless Spectrograph (NIRISS). in *Space telescopes and instrumentation 2012: Optical, infrared, and millimeter wave* (eds. Clampin, M. C., Fazio, G. G., MacEwen, H. A. & Oschmann, Jr., Jacobus M.) vol. 8442 84422R (2012).
7. Stevenson, K. B. *et al.* Transiting Exoplanet Studies and Community Targets for JWST's Early Release Science Program. **128**, 094401 (2016).
8. Bean, J. L. *et al.* The Transiting Exoplanet Community Early Release Science Program for JWST. **130**, 114402 (2018).
9. Fischer, P. D. *et al.* HST Hot-Jupiter Transmission Spectral Survey: Clear Skies for Cool Saturn WASP-39b. **827**, 19 (2016).
10. Lecavelier Des Etangs, A., Pont, F., Vidal-Madjar, A. & Sing, D. Rayleigh scattering in the transit spectrum of HD 189733b. **481**, L83–L86 (2008).
11. Line, M. R. & Parmentier, V. The Influence of Nonuniform Cloud Cover on Transit Transmission Spectra. **820**, 78 (2016).
12. Ackerman, A. S. & Marley, M. S. Precipitating Condensation Clouds in Substellar Atmospheres. **556**, 872–884 (2001).

13. Faedi, F. *et al.* WASP-39b: a highly inflated Saturn-mass planet orbiting a late G-type star. **531**, A40 (2011).
14. Mancini, L. *et al.* The GAPS programme with HARPS-N at TNG. XVI. Measurement of the Rossiter-McLaughlin effect of transiting planetary systems HAT-P-3, HAT-P-12, HAT-P-22, WASP-39, and WASP-60. **613**, A41 (2018).
15. Biazzo, K. *et al.* The GAPS Programme at TNG. XXXV. Fundamental properties of transiting exoplanet host stars. **664**, A161 (2022).
16. Polanski, A. S., Crossfield, I. J. M., Howard, A. W., Isaacson, H. & Rice, M. Chemical Abundances for 25 JWST Exoplanet Host Stars with KeckSpec. *Research Notes of the American Astronomical Society* **6**, 155 (2022).
17. Lodders, K., Palme, H. & Gail, H.-P. Abundances of the Elements in the Solar System. *Landolt Börnstein* **4B**, 712 (2009).
18. Wakeford, H. R. *et al.* The Complete Transmission Spectrum of WASP-39b with a Precise Water Constraint. **155**, 29 (2018).
19. Schneider, A. D. & Bitsch, B. How drifting and evaporating pebbles shape giant planets. II. Volatiles and refractories in atmospheres. **654**, A72 (2021).
20. Hands, T. O. & Helled, R. Super stellar abundances of alkali metals suggest significant migration for hot Jupiters. **509**, 894–902 (2022).
21. Lothringer, J. D. *et al.* A New Window into Planet Formation and Migration: Refractory-to-Volatile Elemental Ratios in Ultra-hot Jupiters. **914**, 12 (2021).
22. Welbanks, L. *et al.* Mass-Metallicity Trends in Transiting Exoplanets from Atmospheric Abundances of $H_2O$, Na, and K. **887**, L20 (2019).
23. Robertson, P., Bender, C., Mahadevan, S., Roy, A. & Ramsey, L. W. Proxima Centauri as a Benchmark for Stellar Activity Indicators in the Near-infrared. **832**, 112 (2016).
24. Changeat, Q. *et al.* Five Key Exoplanet Questions Answered via the Analysis of 25 Hot-Jupiter Atmospheres in Eclipse. **260**, 3 (2022).
25. Lavvas, P., Koskinen, T. & Yelle, R. V. Electron Densities and Alkali Atoms in Exoplanet Atmospheres. **796**, 15 (2014).
26. Parmentier, V., Fortney, J. J., Showman, A. P., Morley, C. & Marley, M. S. Transitions in the Cloud Composition of Hot Jupiters. **828**, 22 (2016).
27. Powell, D. *et al.* Transit Signatures of Inhomogeneous Clouds on Hot Jupiters: Insights from Microphysical Cloud Modeling. **887**, 170 (2019).
28. Roman, M. T. *et al.* Clouds in Three-dimensional Models of Hot Jupiters over a Wide Range of Temperatures. I. Thermal Structures and Broadband Phase-curve Predictions. **908**, 101 (2021).
29. Demory, B.-O. *et al.* Inference of Inhomogeneous Clouds in an Exoplanet Atmosphere. **776**, L25 (2013).
30. MacDonald, R. J. & Madhusudhan, N. HD 209458b in new light: evidence of nitrogen chemistry, patchy clouds and sub-solar water. **469**, 1979–1996 (2017).
31. Benneke, B. *et al.* A sub-Neptune exoplanet with a low-metallicity methane-depleted atmosphere and Mie-scattering clouds. *Nature Astronomy* **3**, 813–821 (2019).

32. Barstow, J. K. Unveiling cloudy exoplanets: the influence of cloud model choices on retrieval solutions. **497**, 4183–4195 (2020).
33. Welbanks, L. & Madhusudhan, N. Aurora: A Generalized Retrieval Framework for Exoplanetary Transmission Spectra. **913**, 114 (2021).
34. Kreidberg, L. *et al.* A Precise Water Abundance Measurement for the Hot Jupiter WASP-43b. **793**, L27 (2014).
35. Pinhas, A., Madhusudhan, N., Gandhi, S. & MacDonald, R. $H_2O$ abundances and cloud properties in ten hot giant exoplanets. **482**, 1485–1498 (2019).
36. Fortney, J. J. *et al.* A Framework for Characterizing the Atmospheres of Low-mass Low-density Transiting Planets. **775**, 80 (2013).
37. Madhusudhan, N., Amin, M. A. & Kennedy, G. M. Toward Chemical Constraints on Hot Jupiter Migration. **794**, L12 (2014).
38. Mordasini, C., van Boekel, R., Mollière, P., Henning, Th. & Benneke, B. The Imprint of Exoplanet Formation History on Observable Present-day Spectra of Hot Jupiters. **832**, 41 (2016).
39. Shibata, S., Helled, R. & Ikoma, M. The origin of the high metallicity of close-in giant exoplanets. Combined effects of resonant and aerodynamic shepherding. **633**, A33 (2020).
40. Öberg, K. I., Murray-Clay, R. & Bergin, E. A. The Effects of Snowlines on C/O in Planetary Atmospheres. **743**, L16 (2011).
41. Eistrup, C., Walsh, C. & van Dishoeck, E. F. Setting the volatile composition of (exo)planet-building material. Does chemical evolution in disk midplanes matter? **595**, A83 (2016).

# Methods

### Data Reduction

We applied six independent data reduction and light curve fitting routines to the data: `nirHiss`, `supreme-SPOON`, `transitspectroscopy`, `NAMELESS`, `iraclis`, and FIREFly. Below, we first describe the major reduction steps taken by each, followed by their light curve fitting methodologies. We note here that in each pipeline the position of the SOSS trace was found to match near-perfectly with that measured during commissioning. Each pipeline therefore used the default wavelength solution for SOSS measured during commissioning and provided in the jwst_niriss_spectrace_0023.fits reference file. We present a summary of all pipelines in Extended Data Table 1.

### The nirHiss Pipeline

`nirHiss` is a Python open-source package that uses the Stage 2 outputs from the `Eureka!` pipeline and performs additional background and cosmic ray removal as well as extraction of the stellar spectra. For this analysis, we took the uncalibrated images and ran our own Stages 1 and 2 calibration using `Eureka!`[42], an open-source package which performs spectral extraction and light curve fitting for several JWST instruments. We use the default steps presented in `Eureka!`, which includes detector-level corrections, production of count rate images, application of physical corrections, and calibrations to individual exposures.

Next `nirHiss` removes background noise sources in a multi-step process. The zodiacal background is first removed by applying the background model provided on the STScI JDox User Documentation website.[1] The background is scaled to a small region of each science integration where there was no contamination from any of the orders; in this case, x∈[190,250], y∈[200,500]. The average scaling—calculated here to be 0.881—is applied to all science integrations. Second, a model of 0$^{th}$ order contaminants is built using the F277W integrations. The F277W integrations were taken after the transit of WASP-39b with the GR700XD/CLEAR pupil element and the F277W filter (throughput centred at $\lambda = 2.776$μm with a bandwidth of $\lambda = 0.715$μm). These observations consist of ten integrations with an exposure time of 49.4s. Observations with the F277W filter contain only the spectral trace of order 1 in the region where $x \leq 460$ pixels, thus allowing for the detection and modelling of 0$^{th}$ order contaminants across the majority of the detector. A median F277W frame is created to identify and mask any bad data quality pixels.

To ensure no additional noise is added from the F277W median frame, we create a 2D background model map using `photutils.Background2D`. To identify regions of the background, we masked the upper-left corner, where the trace is located, and any regions with $> 1.5\sigma$, which includes the 0$^{th}$ order sources. For `photutils.Background2D`, we used a filter size of (3,2) pixels and a box size of (2,2) pixels. Once the background is removed from the median F277W frame, we apply a Gaussian filter with a width of 2 to smooth out any additional small-scale background noise. To apply the median F277W frame to the Stage 2 science integrations, we scaled it to two isolated 0$^{th}$ order sources in the science integrations at $x_1 \in$ [900,1100], $y_1 \in$[150, 250] and $x_2 \in$[1800,2000], $y_2 \in$[150, 250]. We applied the average scaling to all integrations. We found the average F277W background scaling to be 2.81. We apply the scaled background frame to each time-series integration (TSO) integration.

Once the 0$^{th}$ order contaminants are removed, we trace the location of Orders 1 and 2. The spatial profile for NIRISS-SOSS along the column is double-peaked, with a slight dip in the middle. We developed a routine to identify the trace locations using a three-step approach to identifying each order. For each column in the first order trace, we identify the locations of the two peaks, or “ears" and assume the middle of the trace is the median row-pixel between the two “ears". We repeat this process for the third and second orders in that sequence, masking orders once they have been traced. We chose to identify the third order before the second order since it is better spatially resolved and does not overlap with any other orders. The routine creates one main set of traces from a median frame of all observations which is used to extract the stellar spectra. As an additional output, we track the changes in (x,y) pixel positions of each order on the detector across all integrations.

[1] https://jwst-docs.stsci.edu/

After the traces are identified, we continue our reduction to remove any additional noise and cosmic rays/bad pixels. We perform additional $1/f$ noise correction following the routine presented in `transitspectroscopy` (described below). Finally, `nirHiss` identifies and interpolates over cosmic rays. To identify cosmic rays we used the L.A. Cosmic technique wrapped into `ccdproc`[43,44], which identifies pixels based on a variation of the Laplacian edge detection. We identify cosmic rays as pixels with $\sigma > 4$ via this method. We interpolate over any additional bad pixels by taking the median value of the two surrounding pixels along the column. We extract the spectra using a box extraction routine, and ignore any contaminants from overlapping orders or from any potential background orders. We use a box diameter of 24 pixels for both Orders 1 and 2. We do not extract the spectra from Order 3.

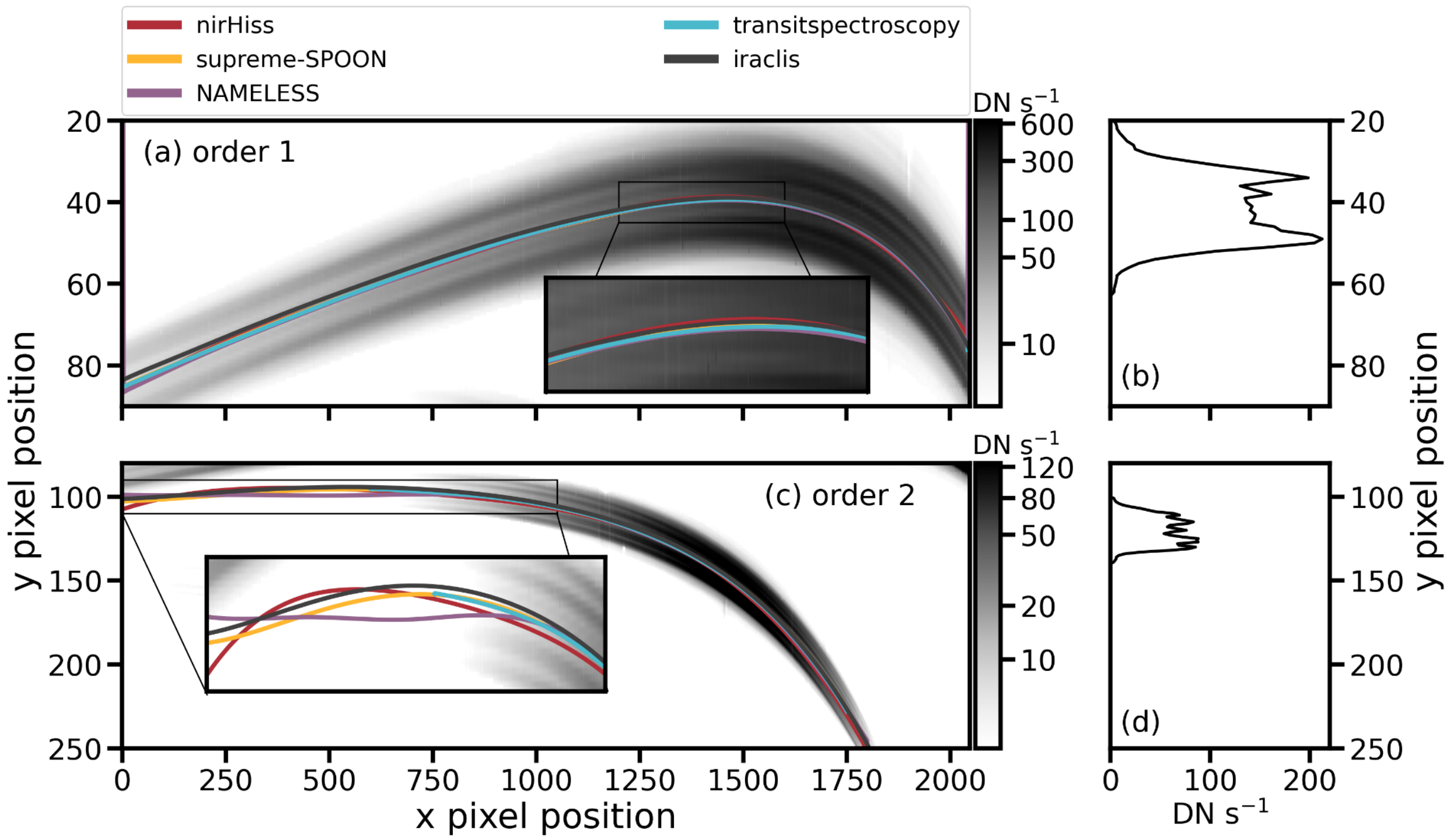

**Extended Data Figure 1: Comparison of (x,y) position of NIRISS Orders 1 and 2 across the detector as modelled from different reduction pipelines. (a/c):** Each pipeline traces the curvature of Orders 1 and 2 using different methods. We show the best-fit trace for Order 1 in the

top row, and Order 2 in the bottom row. We highlight zoomed-in regions to further examine differences. We note that `iraclis` uses the JWST-provided spectral trace. There is generally good agreement between the traces across the entire detector (<1 pixel deviations), with the strongest deviations towards the ends of each trace (e.g., x pixel position < 500 for Orders 1 and 2. **(b/d):** We provide an example spatial profile along the column for Order 1 (b) and Order 2 (d) at x = 1250.
(https://github.com/afeinstein20/wasp39b_niriss_paper/blob/main/scripts/edfigure1.py)

### The supreme-SPOON Pipeline

In parallel, we reduce the WASP-39 b TSO with the independent `supreme-SPOON` (supreme-Steps to Process sOss ObservatioNs) pipeline, which processes SOSS TSOs from the raw, uncalibrated detector images to extracted 1D light curves. An outline of the specific steps is presented below.

For detector-level processing `supreme-SPOON` closely follows Stage 1 of the `jwst` pipeline. All default steps, up to and including the reference pixel correction, are run using their default settings. The reference pixel step is known to provide an inadequate correction of $1/f$ noise for SOSS observations; however, we include it in order to remove group-to-group variations in the bias level, as well as even-odd row variations. At this stage, we remove the zodiacal background from each group. This is accomplished by first calculating a group-wise median frame, and scaling the model background provided in the STScI JDox to the flux level of each group in this median — yielding eight background models, one for each group. The region chosen to calculate the scaling was $x \in [300,500]$, $y \in [210,250]$, where there is minimal contamination from any of the SOSS orders. The $n^{th}$ background model is then subtracted from the corresponding group of each integration.

We then proceed to a more in-depth treatment of $1/f$ noise. Unlike the other pipelines used in this work, `supreme-SPOON` treats $1/f$ noise at the group level instead of the integration level. $1/f$ noise is a time-varying noise source introduced by the voltage amplifiers during the readout of the detector, and therefore the $1/f$ pattern will vary from group-to-group, even within a given integration. To perform the $1/f$ correction, first a median out-of-transit frame is calculated for each group. This group-wise median is then scaled to the flux level of each frame in a given group via the transit curve, and subtracted off — revealing the characteristic $1/f$ striping in the residuals. A column-wise median of this residual map is then subtracted from the original frame. The trace residuals as well as any bad pixels are masked in the median calculation.

From this point we once again proceed with the standard Stage 1 steps of the `jwst` pipeline, with the exception of the `dark_current` step, to obtain the `supreme-SPOON` Stage 1 outputs. The dark current subtraction step is skipped as it was found to re-introduce $1/f$ noise into the data. The dark current level is additionally extremely small (several 10s of electrons/s compared to many thousands for the target signal) and can thus be safely ignored. `supreme-SPOON` only applies the `assign_wcs`, `srctype`, and `flat_field` steps of the Stage 2 `jwst` pipeline to the Stage 1 products. The background subtraction was already performed as part of Stage 1 calibrations. Furthermore, the flux calibration steps (`pathloss` which accounts for light incident on the telescope primary mirror which falls outside of the SUBSTRIP256 subarray, and `photom` which performs the actual photometric flux calibration) are skipped, both because an absolute flux calibration is unnecessary for relative spectrophotometric measurements, and a wavelength-dependent flux calibration is nonsensical for SOSS where contributions from multiple wavelengths from all orders impact a single pixel. At this point, `supreme-SPOON` identifies any remaining hot pixels via median filtering of a median stack of all frames and interpolates them via the median of a surrounding box. These products are the `supreme-SPOON` Stage 2 results.

Stage 3 of the `supreme-SPOON` pipeline is the 1D extraction. This can be performed via two different methods: the first is a simple box aperture extraction on each order, ignoring the order contamination. The second uses `ATOCA` (Algorithm to Treat Order ContAmination)[45] to explicitly model the order contamination. Briefly, `ATOCA` constructs a linear model for each pixel on the detector, including contributions from the first and second diffraction orders, allowing for the decontamination of the SOSS detector — that is, `ATOCA` constructs models of both the first and second orders individually, thereby allowing a box extraction to be performed on each free from the effects of order contamination. Although the effects of this order contamination for differential measurements (e.g., exoplanet atmosphere observations) are predicted to be small (~ 1% of the amplitude of the expected spectral features)[45,46], in the quest to obtain the most accurate possible transmission spectra, this contamination effect is important to take into account. `ATOCA` is currently built into the `extract1dstep` of the official `jwst` pipeline, although it not currently the default option and must be toggled to on via the "soss_atoca" parameter. To improve the performance of `ATOCA`, we do not use the default `specprofile` reference file included in the `jwst` pipeline, but instead construct estimates of the underlying spatial profiles of the first and second order, upon which `ATOCA` relies, using the `APPLESOSS` (A Producer of ProfiLEs for SOSS) algorithm[46]. We determine the centroid positions for each order on a median stack using the "edgetrigger" algorithm[46], and these positions are found to match to within a pixel with the default centroids contained in the jwst_niriss_spectrace_0023.fits reference file; the spectrace file is available on the JWST Calibration Reference Data System (CRDS).[2] The SOSS trace position is furthermore highly stable over the course of this TSO, with RMS variations in x and y positions of ~5 mpix, and RMS rotation of ~0.3$''$. We therefore fix the "soss_transform" parameter to [0, 0, 0], and perform the extraction with a box size of 25 pixels. Any remaining >5σ outliers in the resulting spectra are then identified and clipped. Currently,

[2] https://jwst-crds.stsci.edu/

`supreme-SPOON` does not explicitly treat contamination from zeroth orders of background stars that intersect the trace.

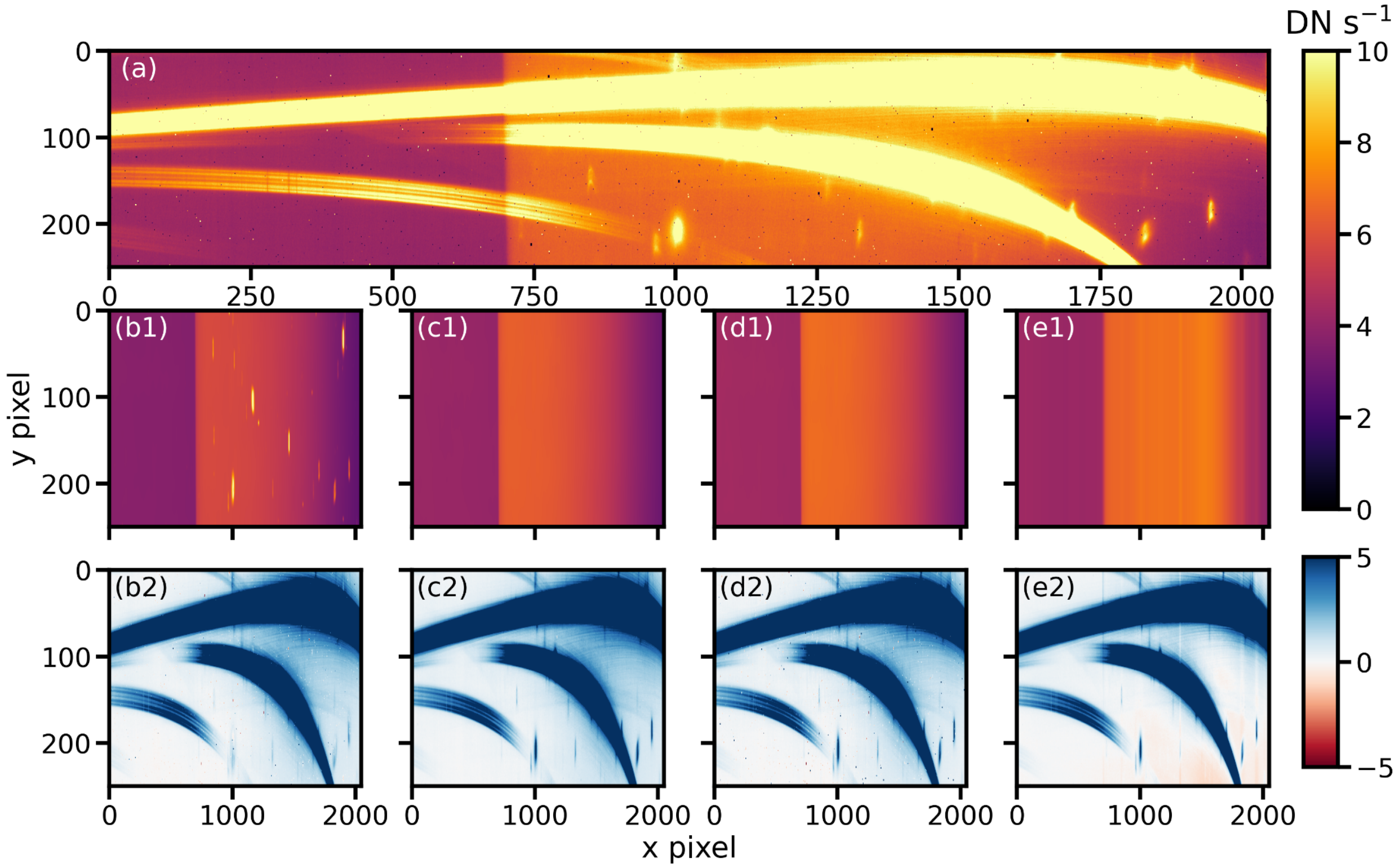

**Extended Data Figure 2: Comparison of averaged background frames computed for each reduction pipeline. Top:** An out-of-transit (OOT) median image from the Stage 2 output files in data numbers per second (DN s$^{-1}$). **Middle:** Estimated median background frames for four example pipelines: (b) `nirHiss`, (c) `supreme-SPOON`, (d) `transitspectroscopy`, and (e) `iraclis`. All reduction pipelines use the predefined zodiacal background model provided on the STScI JDox User Documentation website. `nirHiss` estimates the 0$^{th}$ order contaminants by taking a smoothed median from the F277W filter integrations. We note that the background frame from `supreme-SPOON` (c1) for group eight is shown here, and was scaled by a factor of~0.02 to lie on the same scale as the background from the other three pipelines. `iraclis` subtracts a median per column to remove additional 1/f noise. **Bottom:** Background subtracted median integrations for each pipeline, plotted in [DN s$^{-1}$] but scaled from -5 to 5 to highlight subtle changes in the background. For these integrations, we define OOT as integrations 1-200 and 400-518.
(https://github.com/afeinstein20/wasp39b_niriss_paper/blob/main/scripts/edfigure2.py)

**The transispectroscopy Pipeline**

This third pipeline analysis combines the `jwst` pipeline Stage 1 "rateints.fits" files with `transitspectroscopy`[47]. `transitspectroscopy` completes stellar spectral extraction as well as transit fitting.

The trace positions for NIRISS orders 1 and 2 were determined using `transitspectroscopy.trace_spectrum`. This routine cross-correlates an input function with each column in the detector to find the centre of the different traces via the maximum of the resulting cross-correlation function. In order to follow the shape of the NIRISS order profiles, an input function consisting of a double Gaussian was used with parameters that were trained on the NIRISS/SOSS observations of HAT-P-14 b (JWST Program ID 1541; PI Espinoza): $\mu_2 = -7.5; \sigma_1 = 3.0; \mu_2 = 7.5; \sigma_2 = 3.0$. The trace for order 2 was not fit for pixels $\leq 1040$, as the throughput is not high enough for the method to robustly fit the trace without incorporating nearby contaminants. After identifying the trace positions with this method for both Order 1 and Order 2, both traces are smoothed using a series of spline functions. We find the best-fit parameters for Order 1, which were trained on the HAT-P-14 b observations are: $x_{\mathrm{knots,\ 1}} = [[6, 1200 - 5], [1200, 1500 - 5], [1500, 1700 - 5], [1700, 2041]]$; $n_{\mathrm{knots,1}} = [4,2,3,4]$ and for Order 2 $x_{\mathrm{knots,\ 2}} = [[601, 850 - 5], [850, 1100 - 5], [1100, 1749]]$; $n_{\mathrm{knots,\ 2}} = [2, 2, 5]$.

The zodiacal background was removed by scaling the model background provided on the STScI JDox User Documentation. This model was compared to a small region of the median science integrations where there was little to no contamination from the orders (x∈[500,800], y∈[210,250]). The ratio of all the pixels in this region versus the pixels in the background model was computed, ordered, and the median ratio of all the second quartile pixels was used as the scaling factor between the background model and the data, which was found to be $0.909$. All the integrations had this scaled background subtracted.

Each integration is corrected for $1/f$ noise with the following procedure. First, all the out-of-transit, background-corrected integrations are median combined and scaled by the relative flux decrease produced by the transit event at each integration (i.e., 1.0 for out-of-transit integrations, or about 0.976 for mid-transit). These scaled median frames are then subtracted from each individual integration, which then leaves in the frame only detector-level effects such as $1/f$ noise. We then go column by column and take the median of all pixels in this residual frames within a distance of 20 to 35 pixels from the centre of the trace, and use this as an estimate of the contribution from $1/f$ noise to that given column. This value is then removed from each pixel within 20 pixels from the trace on that column. No correction for Order 1 contamination on Order 2 was made as the contribution is negligibly small in this case[45] — similarly for Order 1 contamination in Order 2 in our extraction.

To extract the resulting background and $1/f$-corrected spectrum, we used the

`transitspectroscopy.spectroscopy.getSimpleSpectrum` routine with a 30-pixel total aperture for both orders. In order to handle obvious outliers in the resulting spectrum due to, e.g., uncorrected cosmic rays and/or deviating pixels, we median-normalised the spectra for each integration and combined them all to form a “master" 1D spectrum for both orders 1 and 2. The median was taken at each wavelength, as well as the error on that median, and this was then used to search for 5σ outliers on each individual integration at each wavelength. If outliers were found, they were replaced by the re-scaled version of this median “master" spectrum.

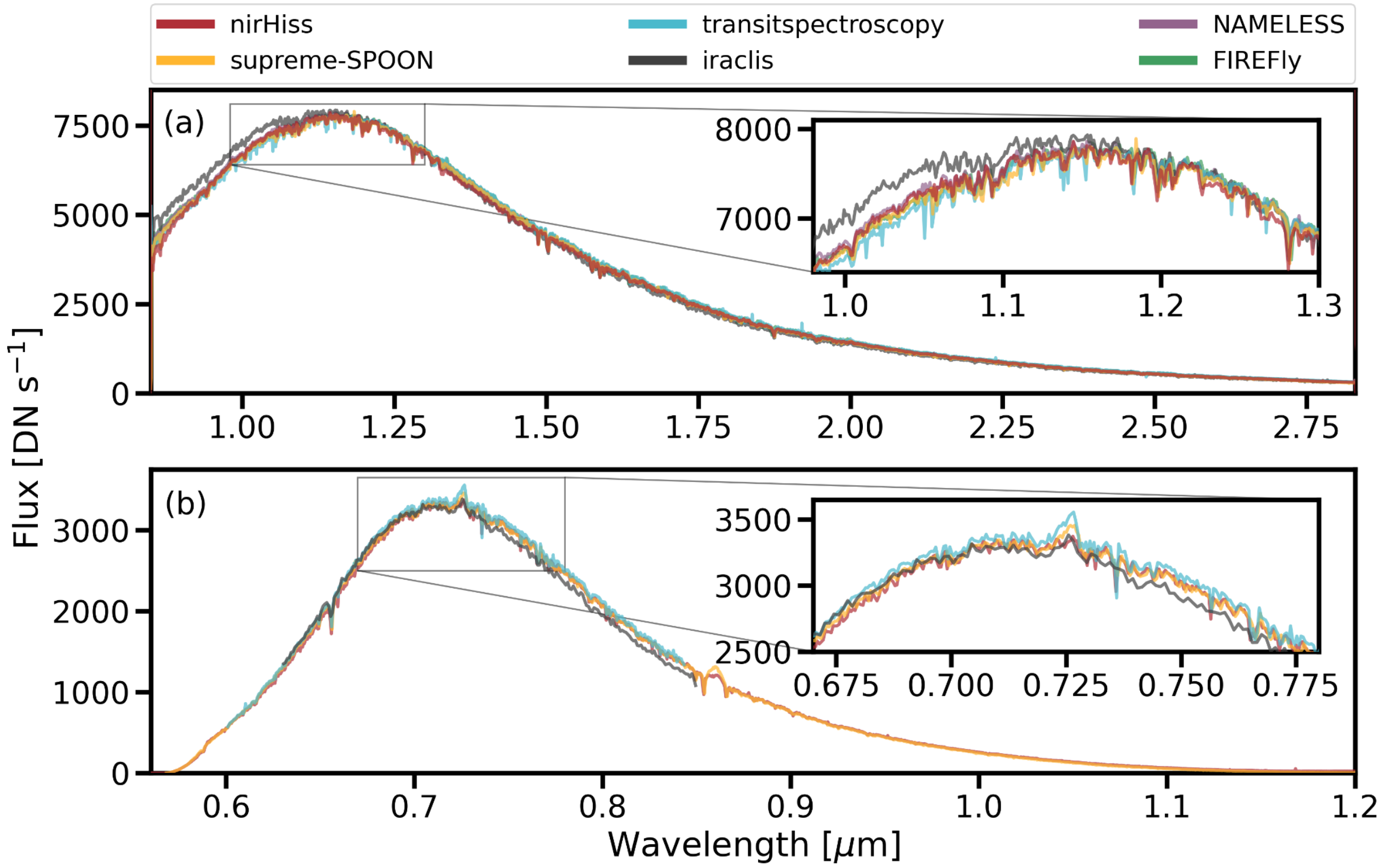

**Extended Data Figure 3: Example extracted stellar spectra from different reduction pipelines.** The inset axes highlight the peak of the spectra. The `supreme-SPOON` spectra are scaled by a factor of 72 to compare the overall shape of the spectra, rather than the extracted flux counts. **Top:** The extracted stellar spectra from Order 1. All pipelines are in relatively good agreement, while the shape of the `iraclis` data changes slightly at λ < 1.1μm. This is likely due to different traces which were used in the spectral extraction routine. **Bottom:** The extracted stellar spectra from Order 2. Differences at λ = 0.725 and 0.86μm are due to differences in removing zeroth order contaminants in the background. The \texttt{iraclis} pipeline does not extract data out past λ > 0.85μm, which is where the order overlap region begins. Across all pipelines, the shape of the spectra, as well as overall absorption features, cosmic ray removal techniques, and noise levels are consistent.
(https://github.com/afeinstein20/wasp39b_niriss_paper/blob/main/scripts/edfigure3.py)

**The NAMELESS Pipeline**

Starting from the `jwst` pipeline Stage 1 products, we use the `NAMELESS` (Niriss dAta reduction MEthod for exopLanEt SpectroScopy) pipeline to go through the `jwst` pipeline Stage 2 with the addition of custom correction routines.

First, we go through the `assign_wcs`, `srctype`, and `flat_field` steps of the `jwst` pipeline Stage 2, opting for a custom background subtraction routine and skipping the `pathloss` and `photom` steps as absolute flux calibration is not needed. After flat-field correction, we scale the model background provided on the STScI JDox User Documentation to a region of the median frame where the contribution from the tail of the three orders is lowest (x $\in$[200,250], y$\in$[400,600]). From the distribution of the scaling values of all pixels within the defined region, we take the 16$^{th}$ percentile as our scaling value and subtract the scaled background frame from all integrations.

We subsequently correct for $1/f$ noise by performing a column-by-column subtraction for each median frame subtracted integrations. The median frame is computed from the out-of-eclipse integrations (integration # $\in$[200,400]) and scaled to each individual integration by dividing the sum of the pixels within the first order by that of the median frame. We then subtract the scaled median frame from all integrations, perform the column-by-column subtraction on the residual frames, and add back the scaled median frame to the corrected residual frames to obtain the $1/f$ corrected integrations.

We detect outliers frame-by-frame using the product of the second derivatives in the column and row directions. This method works particularly well for isolated outliers, as this leads to a strong inflection that corresponds to a large second derivative. Because the spectral orders also lead to larger second derivative values, we divide the frames into windows of 4×4 pixels, compute the local second derivative median and standard deviation, and flag any pixel that is more than four standard deviations away from the median. Furthermore, we also flag pixels with null or negative flux. All identified outliers are set equal to the median value of the window it was identified in.

Finally, we proceed with spectral extraction of the corrected frames by first tracing the sections of the spectral orders that we wish to extract. We trace orders 1 and 2 from $x_1 \in [4,2043]$ and $x_2 \in [4,1830]$ respectively. The centre of the traces is found for each individual column by performing a convolution of the profile with a gaussian filter, where we use the maximum of the convolved profile as the centre of the trace. For the tracing of the second order, we keep the centre of the trace fixed below $x = 900$ as the flux from the first order can bias the tracing method. Furthermore, we smooth the positions of the trace centroids using a spline function with 11 and 7 knots for the first and second order respectively. We perform spectral extraction of the first and second orders at all integrations using the

`transitspectroscopy.spectroscopy.getSimpleSpectrum` routine with an aperture width of 30 pixels.

**The iraclis Pipeline**

We used the `jwst` pipeline Stage 1 “rateints.fits" files with modified routines from `iraclis`[48,49], which was initially designed for HST. The modified routines will be part of the version 2 of `iraclis`, which will become publicly available in the near future. The routines applied to the “rateints.fits" files were flat fielding, bad pixels and cosmic rays correction, sky background subtraction, $1/f$ noise correction, X- and Y-drifts detection, light curve extraction, light curve modelling and planetary spectrum decontamination.

We started our analysis by dividing the images by the appropriate flat field frame (jwst_niriss_flat_0275.fits), as provided by the JWST CRDS. The next step was the bad pixels and cosmic rays correction. As bad pixels we used those with a positive DQ flag in the “rateints.fits" files, excluding the warm pixels, as their large number did not allow for a reliable correction. We also identified extra outliers (cosmic rays or other artefacts) by calculating two flags for each pixel: the difference from the average of the ten horizontally neighbouring pixels (x-flag) and the difference from the average of the ten vertically neighbouring pixels (y-flag). If a pixel's x-flag was 5σ larger than the other pixels in the column and its y-flag 5σ larger than the other pixels in the row, it was identified as a cosmic ray (see also[49]). Both bad pixels and outliers were replaced with the value of a 2D interpolation function, created from the rest of the pixels, similarly to analyses with HST[49].

We then subtracted a column-based sky background frame and a column-based $1/f$ noise frame from each integration. For each image, we first used a trace filter (value $>0.001$ in the jwst_niriss_spectrace_0023.fits, provided by the JWST CRDS), and a column-based 1×median absolute deviation filter to find the illuminated pixels. Then, we calculated the column-based median of the image—using only the unilluminated pixels—and subtracted it from the image. Finally, we calculated the column-based median of the IMFD (Image-MedianFrame Difference)—using only the unilluminated pixels—and subtracted it from the image. This process is not efficient in subtracting 100% of the background contamination, which was removed during the last analysis step (spectrum decontamination).

After reduction, X- and Y-drifts were detected relatively to the first image by comparing the sums along the columns and the rows, respectively, similarly to HST[49]. We then extracted the light curves at high resolution, to avoid the $1/f$ noise. For each spectroscopic image, we initially divided each pixel into a 100×100 grid of subpixels, and, for each subpixel, we calculated the distance from the trace (CD) and the wavelength (λ), creating the $CD_{map}$ and the $\lambda_{map}$, respectively. λ was assigned to each subpixel directly from the wavelength solution (interpolated wavelength solution from the jwst_niriss_wavemap_0013.fits file, provided by the JWST CRDS, shifted by the detected X- and Y-drifts). CD was calculated as the distance between the centre of the subpixel and the point of the trace with the same $\lambda$ (interpolated trace from the jwst_niriss_spectrace_0023.fits file, provided by the JWST CRDS, shifted by the detected X- and Y-drifts). Our high-resolution bins had a $\lambda$-width of 10 Å, ranging between 0.62 and 0.85 μm for order 2 and between 0.85 and 2.8 μm for order 1, and a $CD$-width of 1.5 pixels, ranging from -25 to 25 pixels.

Finally, to construct the light curve of each bin, we applied the following smoothed aperture mask on each spectroscopic image and summed the values of all the subpixels. We chose a smoothed aperture, similarly to HST to reduce the effects of jitter noise:

$$APERTURE = 0.5 \times [erf(\frac{CD_{map} - CD_1}{\sigma_{CD}\sqrt{2}}) - erf(\frac{CD_{map} - CD_2}{\sigma_{CD}\sqrt{2}})] \times [erf(\frac{\lambda_{MAP} - \lambda_1}{\sigma_{\lambda}\sqrt{2}}) - erf(\frac{\lambda_{MAP} - \lambda_2}{\sigma_{\lambda}\sqrt{2}})]$$

where $CD_1$, $CD_2$, and $\sigma_{CD}$ are the bin boundaries and the smoothing factor along the cross-dispersion axis, $\lambda_1$, $\lambda_2$, and $\sigma_\lambda$ are the bin boundaries and the smoothing factor along the dispersion axis. For the smoothing factors we used the values of $\sigma_{CD} = 0.015$ pixels and $\sigma_\lambda = 1$ Å—i.e., ~10% of the bin size. We chose these values for the smoothing factors because lower values would effectively create a sharp-edge aperture, while larger values would force the bins to overlap substantially.

**FIREFly**

While FIREFLy (Fast InfraRed Exoplanet Fitting for Light curves)[50] was written and optimised for reducing NIRSpec-PRISM and G395H time series observations, it worked well on the NIRSISS-SOSS dataset, where it selected and processed the spectrophotometry from the first order only with minimal tuning or intervention. FIREFLy is not written in such a way to extract data from Order 2 ($\lambda < 0.9$μm). In our reduction, we perform standard calibrations on the raw data using the `jwst` Python pipeline, and we perform bad pixel and cosmic ray cleaning on each integration. We perform $1/f$ destriping and background subtraction using a pixel mask generated from the temporally-medianed image that selects regions of the data image below a specified count threshold. We extract the spectrophotometry using an optimised aperture extraction width that is constant in wavelength. The aperture width is selected such that the scatter of the resulting out-of-transit white-light photometry is minimised. We bin the cleaned spectrophotometry in wavelength to create 120 variable-width spectral channels with roughly equal counts in each.

**Light Curve Fitting and Transmission Spectra**

We used a suite of light curve fitting routines to fit the extracted light curves. Each routine fit for orbital parameters from the broadband white-light curves per each order (see Extended Data Figure 4). For the spectroscopic light curves, most routines (`nirHiss/chromatic_fitting`, `supreme-SPOON/juliet`, `transispectroscopy/juliet`, and `NAMELESS/ExoTEP`) fixed the orbital parameters (that is the mid-transit time, $t_0$, semi-major axis to stellar radius ratio $a/R_*$, impact parameter $b$, eccentricity $e$) to the same values to ensure consistency. These parameters were fixed to their best-fitting values from the `transitspectroscopy/juliet` white light curve fit, except for $t_0$ which was fixed to the value obtained from the white light curve in each case. The orbital period of WASP-39 b was fixed to 4.05528 d[9] for all fits. This left the planet-to-star radius ratio $R_p/R_*$, the limb-darkening coefficients, and parameters for any additional systematics models to vary. These four routines also fit spectroscopic light curves at the native instrument resolution. However, two routines, `iraclis` and FIREFly instead fixed the orbital parameters in their spectroscopic fits to values obtained via their white light curve fits. `iraclis` also fit directly for the orbital inclination, $i$, as opposed to $b$ and $a/R_*$ like the other routines. Additionally, iraclis and FIREFly fit for their spectrophotometric light curves at the pixel resolution. We present all of the best-fit white-light curve parameters for Order 1 in Extended Data Table 2. Additionally, for the spectroscopic light curve fits, we only considered the region of order 2 with wavelength <0.85 µm, as the 0.85 – 1.0 µm range is covered at higher signal-to-noise by order 1.

**chromatic_fitting**

`chromatic_fitting` is an open-source Python tool for modelling multi-wavelength photometric light curves. This tool is built on the framework of `chromatic`, a package for importing, visualising and comparing spectroscopic datasets from a variety of sources, including `Eureka!` and the *JWST* pipeline. In this paper we applied `chromatic_fitting` to the `nirHiss` reduction.

`chromatic_fitting` utilises the `PyMC3` (NUTS) sampler[51] to fit the exoplanet transit model to the light curves. First, we fit the white light curves for order 1. The white light curve was generated using an inverse variance weighted average of the unbinned data. We fixed the orbital period to 4.05528 d[9] and assumed a circular orbit. We fit for the mid-transit epoch $t_0$, the stellar mass $M_*$ and radius $R_*$, the impact parameter $b$, the planet-star radius ratio $R_p/R_*$, quadratic limb-darkening coefficients ($u_1$, $u_2$) and out-of-transit flux $F_0$. For the fitted parameters $t_0$, $M_*$, $R_*$, $R_p/R_*$ and $F_0$, we assumed normal priors $N(2459787.56, 0.02^2)$, $N(0.934, 0.056^2)$, $N(0.932, 0.014^2)$, $N(0.146, 0.05^2)$ and $N(1.0, 0.01^2)$ respectively. For $b$ we used a uniform prior between 0 and 1.146, where $b \leq 1 + R_p/R_*$. For the limb-darkening coefficients we calculated the theoretical values from 3D models in `ExoTIC-LD`[52–54] (based on the stellar parameters $T_{eff} = 5512$ K, log$g = 4.47$ dex and Fe/H$= 0.0$ dex[14]) and assumed normal priors around these values with $\sigma = 0.05$. We also included a second-order polynomial in time to describe the systematics with a fixed constant term of 0.0 and normal priors on the first and second-order coefficients $c_1$ and $c_2$ of $N\left(0.0, \left(1e^{-4}\right)^2\right)$. Using PyMC3's NUTS implementation we ran 4000 tuning steps and 4000 draws for the white light curve and the mean parameter values are shown in Extended Data Table 2. We checked for convergence using the rank normalised R-hat diagnostic[55,56].

For each spectroscopic light curve we fixed the period $P$, transit epoch $t_0$, eccentricity $e$, semi-major axis in stellar radii $a/R_*$ and impact parameter $b$ to the white-light parameter values from the transitspectroscopy/juliet routine (Extended Data Table 2). We then fit for the planet-star radius ratio $R_p/R_*$, quadratic limb-darkening coefficients ($u_1$, $u_2$) and out-of-transit flux $F_0$ — for all of these parameters we assumed the same normal priors as for the white light curve. We also included a second-order polynomial in time with the same priors as the white light curve fit. For each wavelength we ran 2000 tuning steps and 2000 draws. The final transmission spectrum was taken as the mean value drawn from the posterior distribution for the planet-star radius ratio with 1σ uncertainties extracted from the 16–84$^{th}$ HDI (highest density interval) region.

**juliet**

We applied the juliet package[57] for light curve fitting to the products of multiple reduction pipelines described above. Here, we give a general overview of the methods and include exact details per fit when appropriate.

For the supreme-SPOON reduced stellar spectra, we fit the white light curve for the mid-transit time, $t_0$, the impact parameter, $b$, the scaled orbital semi-major axis, $a/R_*$, and the scaled planetary radius, $R_p/R_*$; assuming a circular orbit. We also fit two parameters of a quadratic limb-darkening model following the parameterization of[58], as well as an additive scalar jitter and the two parameters of a linear trend with time. We therefore fit seven parameters to the white light curve for each order, using wide, flat priors for each case. We then proceeded to fit the light curves from each individual wavelength bin at the native detector resolution — that is two pixels per bin to roughly account for the extent of the PSF in the spectral direction. This results in 1020 bins for Order 1, and 520 bins for Order 2 as we only consider wavelengths $<$ 0.85 μm. For the spectroscopic fits, we fixed the central transit time to the white light curve value, and the other orbital parameters as described for chromatic_fitting. For the linear trend with time, we used the white light posterior for each of the two parameters as prior distributions for all wavelength bins, whereas for the limb-darkening parameters, we adopted a Gaussian prior centred around the predictions of the ExoTiC-LD package[54] with a width of 0.1. As the SOSS throughput files included with ExoTiC-LD did not cover the full wavelength range of both orders, we instead used the throughputs determined during commissioning and included in the spectrace reference file of the jwst pipeline. We truncated the Gaussian prior at 0 and 1, to prevent the limb-darkening parameters from straying into unphysical regions of the parameter space. We then used flat, uninformative priors for the remaining two parameters, the scaled planetary radius and the scalar jitter. The `supreme-SPOON` white light curve fits have $\chi_\nu^2 = 1.15$ for Order 1, and $\chi_\nu^2 = 1.11$ for Order 2.

For the `transitspectroscopy` reduced stellar spectra, we first fit the white-light light curves of Order 1 and 2 separately. For these, as suggested above, the period was fixed, but all the rest of the parameters were allowed to vary. In particular, we set a normal prior on the time-of-transit centre of $N\left(2459787.5, 0.2^2\right)$ days, where the first value denotes the mean and the second the variance of the prior. A normal prior was also set on $a/R_* \sim N\left(11.37, 0.5^2\right)$ and a truncated normal between 0 and 1 was set for the impact parameter $b \sim TN\left(0.447, 0.1^2\right)$ where the means were set following the work of[59], but the variances are large to account for the variation of these parameters in the literature between different authors. We set a uniform prior for the planet-to-star radius ratio between $0$ and $0.2$, and fixed eccentricity to 0. In addition to those, we fit for a mean out-of-transit offset with a normal prior of $N\left(0, 0.1^2\right)$, and a jitter term added in quadrature to the error bars with a log-uniform prior between 10 and 1000 ppm. To account for systematic trends in the data, we use a Gaussian Process (GP) via `celerite`[60] with a simple Matèrn 3/2 kernel to parametrize those trends. We set log-uniform priors for both the amplitude of this GP from $10^{-5}$ to 1000 ppm, and for the time-scale from $10^{-3}$ days to $0.5$ days. We use the framework of[58] to parameterize limb-darkening via a square-root law, which, following[61], is one of the laws that should give the best results at this level of precision.

For the wavelength-dependent light curves we used a similar setup with two main differences. The first is that we fix the time-of-transit centre, $a/R_*$, and $b$ to their white light values. The second is that we set truncated normal priors on the transformed limb-darkening coefficients $\left(q_1, q_2\right)$ between 0 and 1, with standard deviations of $0.1$ and means obtained via the following method. First, we obtain the non-linear limb-darkening coefficients using an ATLAS stellar model with the closest properties to WASP-39's via the `limb-darkening` software[62]. Then, the square-root law limb-darkening coefficients are obtained following the algorithm of[63], and those are transformed to the $\left(q_1, q_2\right)$ parameterization using the equations in[58]; these are then set as the mean for each wavelength-dependent light curve. We note that we fit the light curves at the pixel level, which means fitting one light curve per detector column. We fit them in parallel using the `transitspectroscopy.transitfitting.fit_lightcurves` routine.

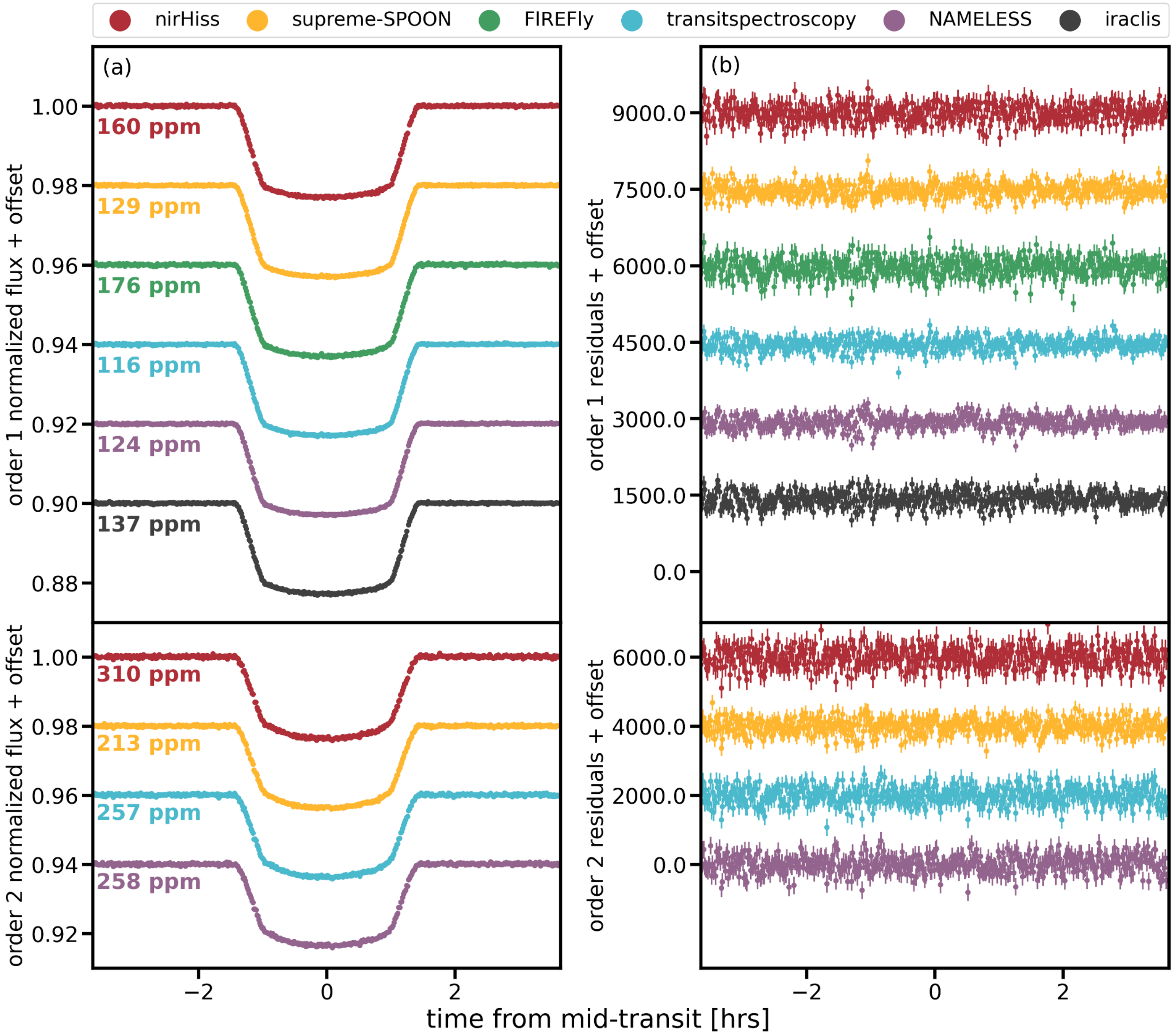

**Extended Data Figure 4: White light curves and residuals between the light curve and best-fit exoplanet transit model per each reduction pipeline. Left:** White light curves for Order 1 (top) and Order 2 (bottom) with the out-of-transit scatter noted in the figure text. **Right:** The residuals, in ppm, between the plotted light curves and best-fit exoplanet transit model. The start of transit ingress and end of transit egress are marked with dashed vertical lines; the transit midpoint is marked with a solid vertical line.
(https://github.com/afeinstein20/wasp39b_niriss_paper/blob/main/scripts/edfigure4.py)

**ExoTEP**

For the `NAMELESS` reduction, we perform light curve fitting on the extracted spectrophotometric observations using the ExoTEP framework[64]. We first fit the white light curves of both orders 1 and 2 separately. We fit for the mid-transit time $t_0$, the planet-to-star radius ratio $R_p/R_*$, and quadratic limb-darkening coefficients ($u_1$, $u_2$)[65,66], while fixing the impact parameter $b$ and semi-major axis $a/R_*$ to the values of the best order 1 white light curve fit from the transitspectroscopy/juliet analysis. We also fit for the scatter σ, as well as a linear systematics model with an offset $c$ and slope $v$. Uniform priors are considered for all parameters. We additionally only discard the first 10 minutes of observations (10 integrations) to remove the exponential ramp. For all light curves, we compute the rolling median for a window size of 11 integrations and bring any data point that is more than four standard deviations away from it to the median value. We fit the light curves using the Markov chain Monte Carlo Ensemble sampler emcee[67] for 1,000 steps using four walkers per free parameter. The first 600 steps, 60% of the total amount, are discarded as burn-in. We then fit the spectroscopic light curves, keeping $t_0$ fixed to its white light value, at a resolution of three pixels per bin for order 1 (680 bins) and one pixel per bin for order 2 from ~0.6–0.9 µm (675 bins). We used 1,000 steps for the spectroscopic fits, once again discarding the first 600 as burn-in.

**iraclis**

We analysed all the light curves using the open source Python package PyLightcurve[68]. For every light curve, PyLightcurve: (I) calculates the limb-darkening coefficients using the ExoTETHyS package[69], the wavelength range of the bin, the response curves for each of the NIRISS orders (jwst_niriss_spectrace_0023.fits file, provided by the JWST CRDS), and the stellar parameters ($T_{eff}$=5540 K, $log g$=4.42 cm/s$^2$, Fe/H=0.14 dex[70]); (II) finds the maximum-likelihood model for the data (an exposure-integrated transit model together with a quadratic trend model using the Nelder-Mead minimisation algorithm included in the SciPy package[71]; (III) removes outliers that deviate from the maximum-likelihood model by more than three times the standard deviation of the normalised residuals; (IV) scales the uncertainties by the RMS of the normalised residuals, to take into account any extra scatter; (V) and, finally, performs an MCMC optimisation process using the emcee package[67]. We initially modelled the first order white light curve (sum of all bins above 0.85 µm with out-of-transit fluxes above 20 DN s$^{-1}$) and fit for the white $R_p/R_*$, the orbital parameters, $a/R_*$ and $i$, and the transit mid-time. We then modelled the spectral light curves, fitting only for the $R_p/R_*$, fixing the orbital parameters, $a/R_*$ and $i$, and the transit mid-time to the above white results. In both cases the models also included a quadratic detrending function that was multiplied by the transit model. After modelling, we applied a spectral decontamination step, taking advantage of the varying total flux across the spectral traces. Due to the contamination we have: $\left(R_p/R_*\right)^2 * (TF - x)/TF$, where $TF$ is the out-of-transit flux (star and contamination) and $x$ is the flux of the contaminating source. Hence, for each wavelength we fitted for $x$ and applied the correction: $\left(R_p/R_*\right)^2_{corr} = \left(R_p/R_*\right)^2 * TF/(TF - x)$. This procedure is effective in removing uniform contamination.The uniform contamination fixes issues of sky background over- or under-correction. It also corrects for order overlap. After the decontamination described above, there was still a contaminating source affecting the spectrum around 0.72 µm, which was not uniform due to the PSF. To separate this source we applied Independent Component Analysis (ICA) on the stellar spectra extracted from various distances from the trace. We used two components to describe the contaminating source and one to describe the stellar spectrum. Finally, we estimated the $\left(R_p/R_*\right)^2$ per wavelength bin using the weighted average of all the bins that had the same wavelength range. We only took into account the bins that had out-of-transit fluxes above 20 DN $s^{-1}$. This choice effectively applied an optimal aperture size for each wavelength bin.

**FIREFly**

To extract the transmission spectrum, we fit for the planet's transit depth at each wavelength using a joint light curve and systematics model. We use the orbital parameters recovered from an MCMC fit to the white light curve, and fix them at each wavelength channel for our fit. We fit for the two quadratic limb-darkening terms $a$ and $b$ at each wavelength channel. We find that the best-fit limb-darkening coefficients are uniquely determined, and deviate by a constant offset relative to model coefficients. Our fits are performed iteratively using the Python package `lmfit`. The light curves show a typical photometric scatter of 0.3% per integration, and the typical transit depth uncertainties vary between 150–300 ppm, which is in line with near photon-limited precision. More details of the FIREFly fitting routine can be found in[50].

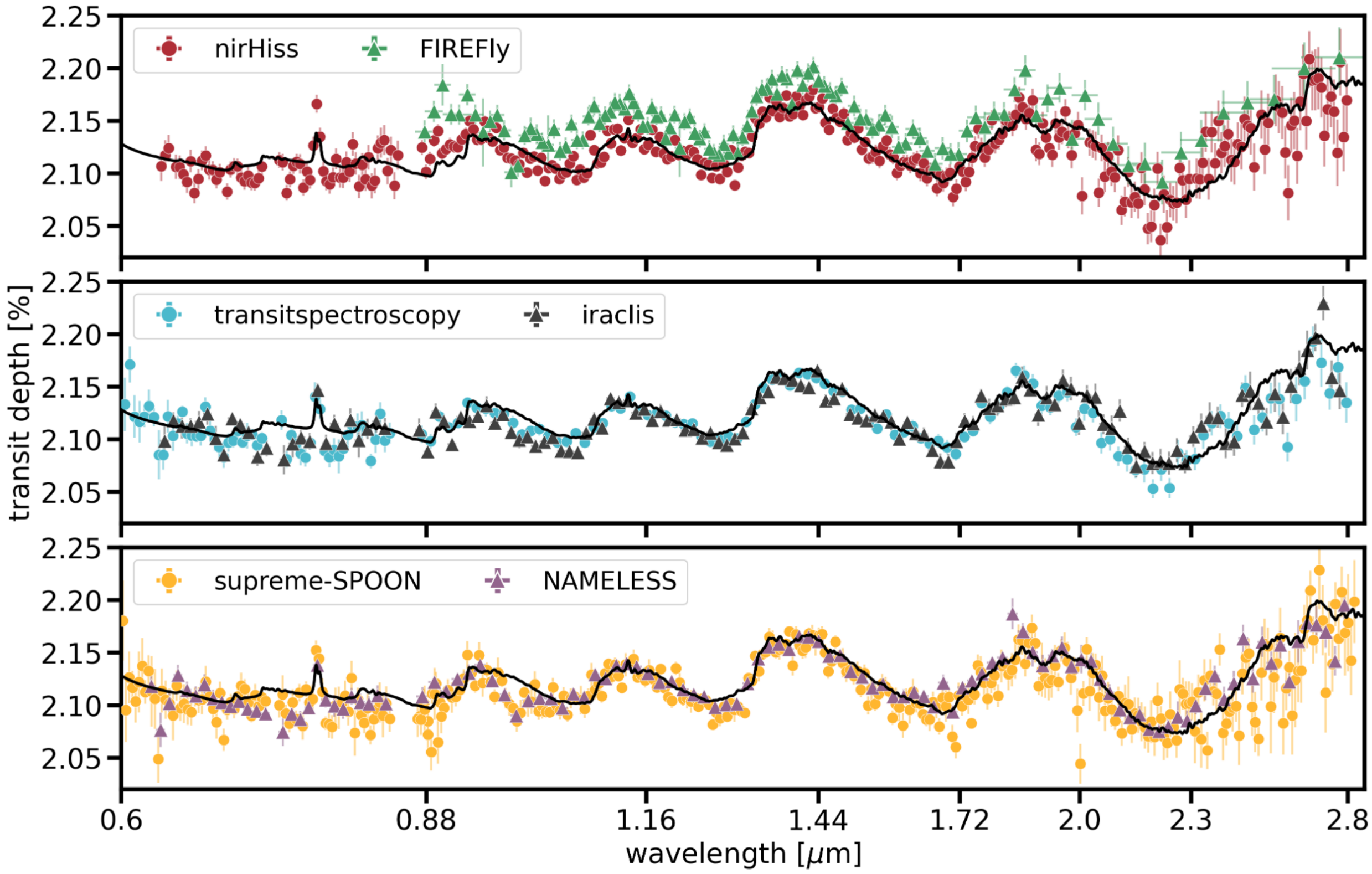

**Extended Data Figure 5: Transmission Spectra for WASP-39b for all reduction techniques.** Our best-fit reference model to the `nirHiss` spectrum (red) is plotted as a black solid line on both panels and the spectra are separated into three panels for ease of reading. Wavelengths which overlap with $0^{th}$ order contaminants are masked.
(https://github.com/afeinstein20/wasp39b_niriss_paper/blob/main/scripts/edfigure5.py)

## Atmospheric Models

To interpret the measured transmission spectrum, we performed an extensive comparison with grids of synthetic transmission spectra. We tested several independent atmospheric models to avoid any model-dependent interpretation of the data. Unless otherwise noted, all of our grids have assumed radiative-convective-thermochemical equilibrium to estimate atmospheric compositions. The exploration of atmospheric models with fewer assumptions (e.g., without the assumption of chemical equilibrium with metallicity and C/O as free parameters) and those considering other effects of disequilibrium chemistry is left for future work.

We derive basic interpretations for the observed spectrum based on four independent model grids, `ATMO`, `PHOENIX`, `PICASO`, `ScCHIMERA` . Each grid contains pre-computed transmission spectra at various atmospheric properties, such as metallicity [M/H], carbon-to-oxygen C/O ratio, and cloud properties, using from gray to Mie-scattering cloud opacity (see next subsection for details). The `ScCHIMERA` grid considers additional model advancements 1) various cloud treatments, including a gray cloud, gray + power-law cloud opacity, and physically motivated (i.e., droplet sedimentation) cloud model, 2) the impact of inhomogeneous cloud coverage along the planetary terminator, and 3) the K/O ratio as a grid dimension. `ScCHIMERA` provides the best fit to the observations compared to the other three grids and informs the results presented in the main text.

## Grid Search with Pre-Computed Forward Models

Here, we introduce the independent grids of pre-computed transmission spectra, their model description, and the main results from these grid fits. We first present the three grids that do assume horizontally homogeneous clouds.

### ATMO

The atmospheric PT profile is computed using the 1D radiative-convective equilibrium model `ATMO`[72–75]. The model includes the molecular/atomic opacity of $CH_4$, CO, $CO_2$, $C_2H_2$, Cs, FeH, HCN, $H_2O$, $H_2S$, K, Li, Na, $NH_3$, $PH_3$, Rb, $SO_2$, TiO, and VO, for which the adopted line list is summarised in[75]. The line lists of several key species are: $H_2O$[76], $CH_4$[77], $CO_2$[78], CO from HITEMP2010[79], K from VALD3[80]. We considered atmospheric metallicities of [M/H]=$-1.0$, $+0.0$, $+1.0$, $+1.7$, $+2.0$, $+2.3$, C/O ratios of C/O=$0.35$, $0.55$, $0.7$, $0.75$, $1.0$, $1.5$, planetary intrinsic temperatures of $T_{\rm int}$=100, 200, 300, 400K, and day-night energy redistribution factors of $0.25$, $0.50$, $0.75$, $1.0$, where full heat redistribution corresponds to $0.5$. The cloudy models include small particle opacity as the Rayleigh scattering gas opacity enhanced by a factor of either 0 or 10, while large particle opacity is equated to the $H_2$ Rayleigh scattering opacity at 0.35μm enhanced by a factor of 0.5, 1.0, 5.0, 10.0, 30.0, and 50.0. In total, the ATMO grid consists of 484 cloud-free and 6292 cloudy atmosphere models. We only consider horizontally homogeneous clouds in the ATMO grid fits.

**PHOENIX**

The atmospheric PT profile is computed using the 1D radiative-convective equilibrium model PHOENIX[81–83]. We considered atmospheric metallicities of [M/H]=− 1.0, + 0.0, + 1.0, + 2.0; C/O ratios ranging from C/O=0.3 to 1 divided into 136 grid points, planetary intrinsic temperatures of $T_{\rm int}$ = 200 and 400K, and day-night energy redistribution factors of 0.172, 0.25, and 0.351, where full heat redistribution corresponds to 0.25. The model includes various chemical species: CH, $CH_4$, CN, CO, $CO_2$, COF, $C_2$, $C_2H_2$, $C_2H_4$, $C_2H_6$, CaH, CrH, FeH, HCN, HCl, HF, HI, HDO, $HO_2$, $H_2$, $H_2S$, $H_2O$, $H_2O_2$, $H_3$+, MgH, NH, $NH_3$, NO, $N_2$, $N_2O$, OH, $O_2$, $O_3$, $PH_3$, $SF_6$, SiH, SiO, $SiO_2$, TiH, TiO, VO, and atoms up to U. The line list of $H_2O$ is from BT2[76], other molecular lines from HITRAN 2008[84], and atomic lines from the database of Kurucz[85]. For cloudy models, the small non-gray cloud particle opacity is treated as a sum of Rayleigh scattering opacity of all gas species enhanced by a factor of either 0 (clear atmosphere) or 10; large gray particle opacity is treated as gray cloud deck pressure levels of 0.3, 3, and 10 mbar. In total, the PHOENIX grid consists of 95 cloud-free and 380 cloudy atmosphere models. We only consider horizontally homogeneous clouds in the PHOENIX grid fits.

**PICASO 3.0**

Similarly to the grids of models presented above, we precomputed atmospheric pressure-temperature (PT) profiles using the 1D radiative-convective equilibrium model PICASO 3.0[86–89] for the atmospheric metallicity of [M/H]=− 1.0, − 0.5, + 0.0, + 0.5, + 1.0, + 1.5, + 1.7, + 2.0; atmospheric bulk C/O ratios of C/O = 0.229, 0.458, 0.687, 1.1; planetary intrinsic temperatures of $T_{\rm int}$ =100, 200, 300K; and heat redistribution factors of 0.5 and 0.4, where full heat redistribution corresponds to 0.5. The model includes 29 chemical species: $CH_4$, CO, $CO_2$, $C_2H_2$, $C_2H_4$, $C_2H_6$, CrH, Cs, Fe, FeH, HCN, $H_2$, $H_2O$, $H_2S$, $H_3$+, OCS, K, Li, LiCl, LiH, MgH, $NH_3$, $N_2$, Na, $PH_3$, Rb, SiO, TiO, and VO. The line lists of several key species are: $H_2O$[90], $CH_4$[91], $CO_2$[92], CO[93], K from VALD3[80]. For cloudy models, we post-processed the computed PT profiles using the droplet sedimentation model Virga[94,95] that determines the vertical distributions of cloud mass mixing ratio and mean particle size from the balance between downward mass flux of gravitational sedimentation and upward mass flux of eddy diffusion. We vary vertically-constant eddy diffusion coefficients of $K_{zz}$ = $10^5$, $10^7$, $10^9$, $10^{11}$ and vertically-constant sedimentation parameters of $f_{sed}$ =0.6, 1.0, 3.0, 6.0, 10.0. The $f_{\rm sed}$ value is defined as the ratio of the mass-averaged sedimentation velocity of cloud particles to the mean upward velocity of the atmosphere, and a smaller $f_{\rm sed}$ yielding more vertically extended clouds[95] see e.g.,[96]. We have assumed horizontally homogeneous clouds and accounted for the formation of $MgSiO_3$, MnS, and $Na_2S$ clouds. Then, we postprocessed the atmospheric properties to compute synthetic transmission spectra. We note that the optical properties of the flux-balanced cloud is computed by the Mie theory[97] under the assumption of a log-normal particle size distribution with a mean particle size translated from $f_{sed}$[94]. In total, the PICASO grid consists of 192 cloud-free and 3840 cloudy atmosphere models.

We compare the NIRISS SOSS spectrum (binned to R=300) to each of these model grids and summarise the best fits in the top panel of Extended Data Fig. 6. For each cloudy and clear model we tested, we compute $\chi^2/N_{obs} = 2.98 - 8.55$ between the data and the models, with specific values per model indicated in the legend of Extended Data Fig. 6. All of our forward model grids consistently indicate super-solar metallicity ([M/H]=1–2) and sub-solar C/O ratio. Each best-fit spectrum shows different structures at >2μm, as the spectra at these wavelengths are more sensitive to the treatment of cloud properties (see next subsection for details). The best-fit spectra from PICASO, ATMO, and Phoenix indicate atmospheric metallicities of [M/H]=1.7, 1.0, and 2.0, respectively. These models also consistently indicate C/O ratios between $0.229$–$0.389$, corresponding to the lowest C/O ratio grid point in each grid (see the main text for why models prefer lower C/O ratios). Thus, the super-solar metallicity and sub-solar C/O ratio of WASP-39b are consistent across the different model interpretations of the NIRISS-SOSS transmission spectrum.

We also find that clear atmospheric models fail to fit the observed spectrum even at very high metallicity ([M/H]=2.0), as shown in the bottom panel of Extended Data Fig. 6. The clear models fail to match the amplitudes of $H_2O$ absorption features at $\lambda = 0.9$, $1.15$, $1.4$, and 1.8μm simultaneously. The clear ATMO models fit the data better than the clear PICASO and Phoenix models because the ATMO grid allows lower heat redistribution factors (i.e., cooler atmosphere). The clear models also overestimate the transit depth at $\lambda \sim 2$ μm because of a strong $CO_2$ absorption resulting from the inferred high metallicity ([M/H]=2.0). The inability of clear atmosphere models to fit the overall NIRISS spectrum strongly indicates the presence of clouds in the atmosphere, and emphasises the ability of the NIRISS wavelength coverage to break the cloud property-metallicity degeneracy. The best-fit cloud properties are $f_{sed}=1$ and $K_{zz} = 10^9$ cm$^2$ s$^{-1}$ for Virga clouds in PICASO, a gray cloud opacity of $5\times$ the $H_2$ Rayleigh scattering opacity at 0.35μm for ATMO, and a gray cloud top pressure of $3 \times 10^{-4}$ bar for Phoenix.

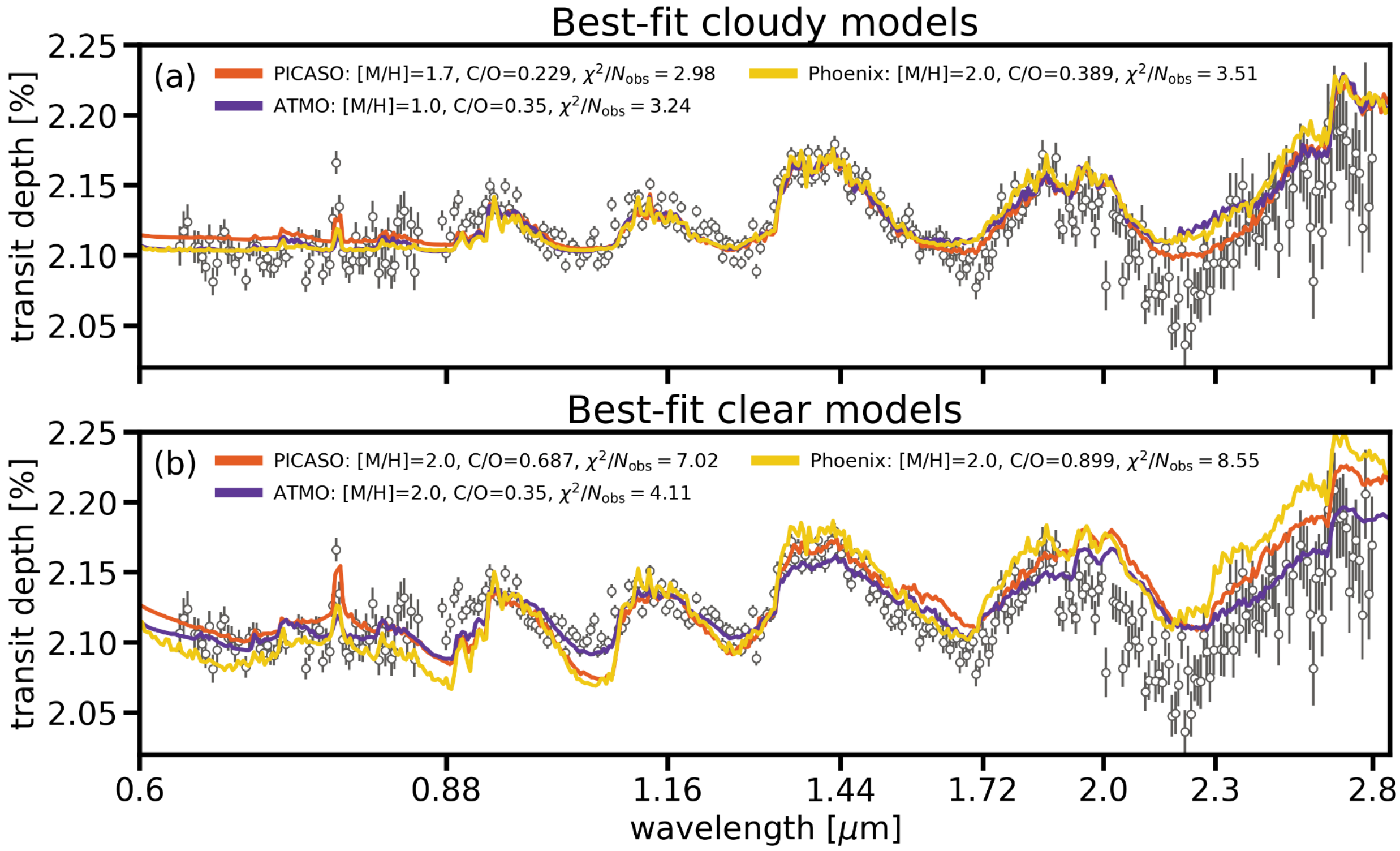

**Extended Data Figure 6: A summary of pre-computed forward model fits to NIRISS-SOSS spectrum. Top:** Each coloured line shows the best-fit spectrum from the `PICASO`, `ATMO`, and `PHOENIX` cloudy grid. The chi-squared values per number of data points ($N_{obs}$ = 327) are $^2/N_{obs}$= 2.98, 3.24, and 3.51 for the `PICASO`, `ATMO`, and `PHOENIX` grids, respectively. All grid models consistently indicate a super-solar metallicity of [M/H] = 1– 2 and a sub-solar C/O ratio. **Bottom:** The same as the top panel, but for the best-fit clear atmosphere models. The clear models yield noticeably worse fits to the data: $^2/N_{obs}$ = 7.02, 4.11, and 8.55 for the `PICASO`, `ATMO`, and `PHOENIX` grids, respectively, which strongly indicates the presence of clouds in the atmosphere.
(https://github.com/afeinstein20/wasp39b_niriss_paper/blob/main/scripts/edfigure6.py)

**Grid Search with ScCHIMERA**

The NIRISS transmission spectrum offers key insights into the atmospheric properties of WASP-39b over a broad wavelength range. The simultaneous detection of $H_2O$ and K, alongside possible indications of carbon-bearing species, allows us to explore equilibrium models for which the potassium-to-oxygen (K/O) ratio is an additional dimension besides the commonly employed C/O and metallicity parameters. Furthermore, as explained in the previous subsection (see also Fig. 3 demonstrating how clouds contribute to the NIRISS spectrum), the broad wavelength coverage of these NIRISS observations makes it possible to explore more complex cloud models beyond traditional gray and homogeneous cloud models. To explore these considerations, we implement the `ScCHIMERA` grid as explained below.

**ScCIMERA**

Previous implementations of this framework include[98–100], where the methods are described in detail. Implementations of this procedure to JWST data include ref. [101] For a given set of planetary parameters, our methods pre-compute the temperature-pressure structure of the planetary atmosphere and the thermochemical equilibrium gas mixing ratio profiles. The computations are performed on a grid of atmospheric metallicity ([M/H], e.g., $log_{10}$ enrichment relative to solar[17]) spaced at 0.125 dex values between 0 and 2.25 (e.g., 1 to 177×solar) and carbon-to-oxygen (C/O) ratios at values of 0.2, 0.35, 0.45, 0.55, 0.65, 0.7, and 0.8. Unlike previous implementations of this framework, and to better understand the NIRISS-SOSS observations presented, we include a dimension to our grid exploring the potassium-to-oxygen ratio ([K/O], i.e., $log_{10}$ enrichment relative to solar[17]) with spacing of 0.5 dex between -1 and 0, and 0.1 dex between 0 and 1, overall spanning a range from -1 to 1, or 0.1 to 10×solar. In these calculations, the atmospheric metallicity (M/H) scales the sum of K, C and O. This sum determines the final elemental abundances after scaling M/H, C/O and K/O. That is, the total oxygen elemental abundance is $O' = \frac{M/H}{K/O+C/O+1}$, the total carbon elemental abundance is $C' = O' \times C/O$, and the total potassium elemental abundance is $K' = O' \times K/O$. Additionally, we explore the energy redistribution (f) between the day and night sides of the planet[102], with values of 0.657, 0.721, 0.791, 0.865, 1.0 , 1.03 , 1.12 , 1.217, 1.319 in our grid, where $f = 1.0$ and $2.0$ correspond to full day-to-night heat redistribution and dayside only redistribution, respectively.

The transmission spectrum of the planet is computed with CHIMERA[96,103,104] using the converged atmospheric structures. We compare the observations to these models in a Bayesian inference framework using the nested sampling algorithm `MultiNest`[105] through its python implementation `PyMultiNest`[106], and obtain an optimal [M/H], C/O, [K/O], and f via nearest neighbour search in the grid. When computing the transmission spectrum for a given set of ([M/H], C/O, [K/O], f), we additionally adjust the 1 bar planetary radius controlling the absolute transit depth (an arbitrary pressure with no direct impact on the inferred properties, see e.g.,[4]), and model different cloud treatments. The opacity sources considered are $H_2$-$H_2$ and $H_2$-He CIA[107], $H_2O$[90,108], $CO_2$[108], CO[79], $CH_4$[79], $H_2S$[109], HCN[110], Na[111,112], and K[111,113], and were computed following the methods described in[114,115]. The cloud models considered are 1) a basic cloud model with a gray, uniformly vertically distributed cloud opacity ($\kappa_{cloud}$); 2) a gray+power-law cloud model that accounts for non-gray opacity of small-size particles as a vertically uniform power-law opacity (i.e., a parameter for the scattering slope and and a Rayleigh-enhancement factor, e.g,[30,33,116,117]) in addition to gray cloud component, which is expressed by gray cloud-deck of infinite opacity at a given atmospheric pressure; and 3) a droplet sedimentation model[94] (assuming enstatite grains) where parameters capture the eddy diffusion coefficient and the ratio of sedimentation velocity to characteristic vertical mixing velocity (see also the description of `PICASO` above). For cloud treatments 2 and 3, we also consider the possibility of inhomogeneities around the planetary limb by considering a linear combination of clear and cloudy models[e.g. 11], key for breaking degeneracies between metallicity and cloud properties (e.g.,[2,4]). We assume the same PT profile for both cloudy and clear limbs in the inhomogeneous cloud models and leave an investigation on the possibility of different PT profiles on those regions to future studies.

**Identification of Absorbers and Model Selection**

We perform our Bayesian inference using all model combinations with the `ScCHIMERA` grid on four different data resolutions for the `nirHiss` transmission spectrum: R=100, R=300, native instrument resolution ($R_{order\ 1}$=910; $R_{order\ 2}$=830), and pixel-level resolution ($R_{order\ 1}$=1820; $R_{order\ 2}$=1660). Resolutions are given at the reference wavelengths of $\lambda = 1.791$μm for Order 1 and $0.744$μm for Order 2. We test the robustness of our inferences against different binning and convolution strategies and find the results, i.e., the bulk atmospheric properties [M/H], C/O, and K/O, consistent regardless of the resolution of the data. We find a fiducial combination of parameters that can best explain the spectrum (that we call the reference model) with full redistribution (f=1, matching predictions that planets in this temperature regime are unlikely to possess strong day-to-night temperature contrast[118–120]), [M/H]=1.375 (i.e., $\sim 20 \times$ solar), C/O=0.2, and [K/O]=0.1. With these atmospheric properties, the data are best explained by the droplet sedimentation model (`ScCHIMERA` cloud model 3) and inhomogeneous cover. However, the gray+power-law model (`ScCHIMERA` cloud model 2) with inhomogeneous cover provides a comparable fit to the data. Using R=300 data, the homogeneous droplet sedimentation model (model 3) is preferred over homogeneous gray cloud (model 1) at $\gtrsim$8σ, which strongly indicates the non-gray nature of cloud opacity. Meanwhile, the inhomogeneous droplet sedimentation model is preferred over the homogeneous droplet sedimentation model cloud at $5\sigma$. This is evidence that for the same model 3 inhomogeneous cloud coverage is preferred.

We explore the contribution of different chemical species to our reference model by performing the Bayesian inference using the inhomogeneous cloud model 3 and artificially disabling the contribution of a selected chemical species, one at a time. By redoing the Bayesian inference, we are able to compare the Bayesian evidence by computing the Bayes factor and converting to a 'sigma' detection significance using the prescription in[2] (see also[22]). We detected $H_2O$ at $> 30\sigma$, K at $6.8\sigma$, CO at $3.6\sigma$, and no significant detections of Na, $CH_4$, $CO_2$, HCN, and $H_2S$. The best-fit metallicity across all models is $\sim$ 10–30 $\times$ solar, the best fit K/O ratio 1–2 $\times$ solar, and C/O ratio 0.2. Taking the average and standard deviation of the best-fit results for all 20 runs (i.e., 5 models on 4 data resolutions) we find an average M/H=19$\times$ solar with a standard deviation of 5$\times$ solar and an average K/O=1.5$\times$ solar with a standard deviation of $0.26 \times$ solar.

**Wavelength Sensitivity to Inferences**

We investigate the dependence of the inferred atmospheric properties on the spectral range of the observations by performing the same Bayesian inferences described above on the spectrum blue-ward of 2 μm (see panel b of Extended Data Fig. 7). This exercise is repeated on all 20 models-data-combinations from ScCHIMERA . With the exception of the solar-to-super-solar K/O ratio, inferences about the atmospheric metallicity, C/O ratio, and clouds are primarily driven by the shallower transit depth seen between $\lambda = 2.1 - 2.3$μm. This wavelength region is where the traces of Orders 1 and 2 overlap on the detector. To assess the robustness of our results, we explore different data treatments that could affect the final spectrum. First, we find there are no zeroth order background contaminants that could be diluting the transit depth in this region. Second, we extract the transmission spectra and fit for dilution between the orders (supreme-SPOON data reduction) and without accounting for the overlap (supreme-SPOON, nirHiss and transitspectroscopy). The evidence for minimal dilution stems from reducing the data through both methods with the same pipeline (supreme-SPOON), which uses the same steps for the entire reduction process along the way, with the exception of fitting and not fitting for dilution. Both techniques yield similar results between $\lambda = 2.1 - 2.3\,\mu m$. We note that the contamination from Order 2 into Order 1 was previously shown to be between 8 – 12 ppm[45] and is therefore negligible.

We find that without the data red-wards of 2 μm, the [M/H] value is more scattered across models and resolutions with an average metallicity of 61×solar for the 20 runs and a standard deviation of $28\times$ solar. On the other hand, the inference on the C/O ratio remains consistently 0.2 across all models and resolutions. Similarly, the K/O ratio remains solar-to-super-solar with an average of 1.89× solar and a standard deviation of 0.29× solar.

These results confirm the necessity for the NIRISS broad wavelength coverage to constrain the atmospheric metallicity of a planet[2,4,11]. Without the transit depth decrease at $2.1$ μm, our models do not exhibit a preference for cloud models 2 and 3 over cloud model 1, nor do they prefer the presence of inhomogeneities in the cloud cover. Without these constraints on the cloud properties, a wide range of metallicities can provide an equally good fit to the observations blue-wards of 2 μm when combined with different cloud properties, preventing reliable constraints on the metallicity.

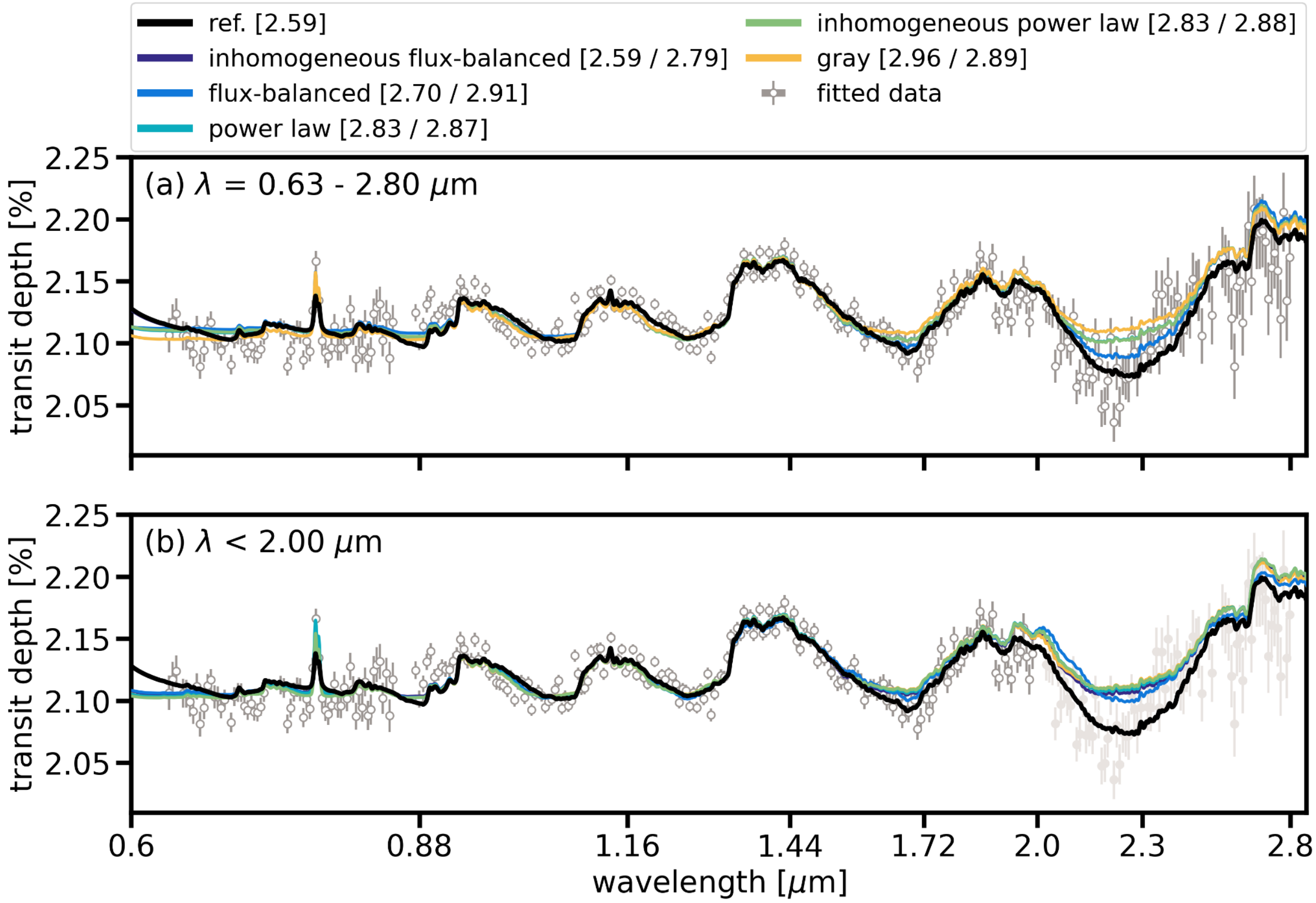

**Extended Data Figure 7: A demonstration of how the redder wavelength coverage of NIRISS-SOSS drives the inference on cloud structure for WASP-39b.** We fit the NIRISS-SOSS spectrum (gray) using a suite of cloud models to derive the best-fit C/O and metallicity. Here, we demonstrate how the best-fit model for each cloud treatment changes as a function of what wavelength region we fit. **Top:** The best-fit models when using the entire wavelength coverage of NIRISS-SOSS. **Bottom:** The best-fit models when using λ < 2 μm, which excludes the overlapping region between orders on the detector. The reference spectrum (black) on both panels corresponds to the best-fit inhomogeneous droplet sedimentation model for the entire wavelength coverage. The fitted data are presented as dark grey points. The quoted numbers in brackets in the legend are the respective $^2$/N for each fit for the top (left value) and bottom (right value). The difference between cloud models is within the noise of the NIRISS-SOSS data when fitting to λ < 2μm. It is clear that fitting the entire NIRISS-SOSS wavelength coverage results in a lower $^2$/N and better fit.
(https://github.com/afeinstein20/wasp39b_niriss_paper/blob/main/scripts/edfigure7.py)

The exploration of these models is summarised in Extended Data Fig. 7. The top panel shows the different cloud treatments and their goodness of fit to the data. Overall, models with inhomogeneous cloud cover best explain the data, with the flux-balanced cloud of model 3 giving the lowest $\chi^2$. The bottom panel contrasts the reference model against the results from all cloud models when using data blue-wards of 2 μm only. Without the information contained in the dip in transit depth at 2.1 μm, all cloud treatments provide an equally good fit and overestimate the transit depth between 2.0 μm and 2.3 μm.

**K/O Inferences**

We explore the possibility of constraining the potassium-to-oxygen ratio using NIRISS-SOSS. As explained above, across different models and data resolutions, our results suggest that the observations of WASP-39b are best explained by a solar-to-super-solar K/O ratio. To further explore this, we repeat our Bayesian inference for all 20 model-data configurations (5 models × 4 resolutions) using the observations blue-wards of 0.8 μm. From high-resolution to low-resolution observations and for all cloud model configurations, we find that all 20 runs prefer models with solar or super-solar K/O ratios for WASP-39b ranging from 1 to 10× solar. The average across the 20 runs is 2.12× solar and a standard deviation of $2.33$ × solar, with the relatively larger standard deviation resulting from two inferences of highly super solar K/O ratios of 7× solar or greater for observations at the pixel-level resolution.

Using the reference model atmospheric properties, (e.g., [M/H]=1.37, C/O=0.2, full redistribution f=1), we search for the best-fit [K/O] while simultaneously adjusting the 1 bar radius and the parameters for the inhomogeneous cloud model 3, when only using the observations blue-wards of 0.8 μm. The best-fit [K/O]=0.4 is consistent with the inferences using all the data and the data blue-wards of 2.0 μm only. This model is shown in Extended Data Figure 8 in green. For the best fit cloud parameters and 1 bar radius, we compute a series of K/O ratios spanning sub-solar and super-solar values. Our results find that sub-solar K/O ratios are disfavoured to 2σ, while super-solar values $\gtrsim 0.7$ are disfavoured to 5σ.

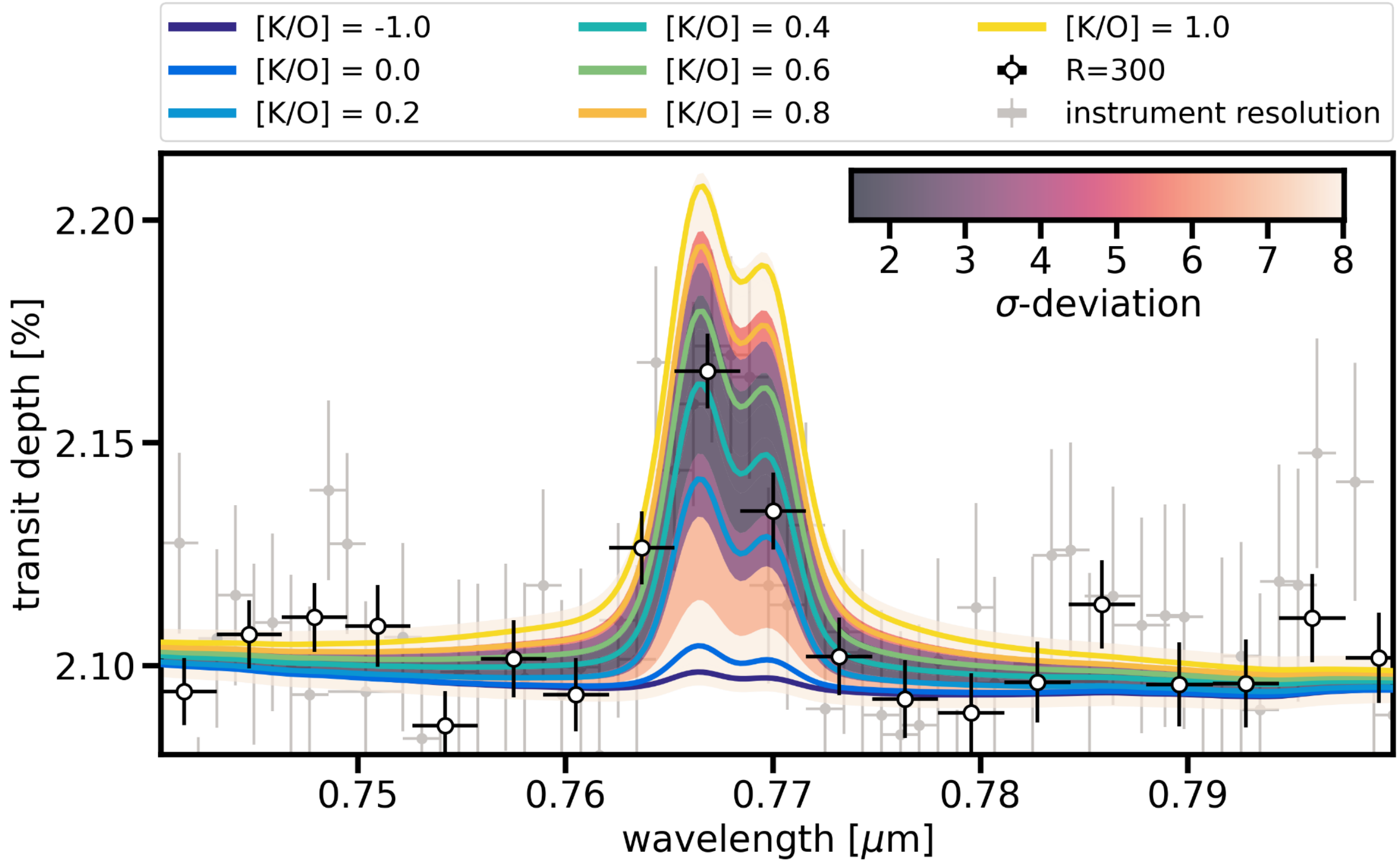

**Extended Data Figure 8: Evidence for super-solar [K/O] in WASP-39b.** We fit for [K/O] while keeping the rest of the model parameters (e.g., C/O, metallicity, and redistribution) the same as our reference model and fitting for the cloud parameters and scaled planetary radius. Here, we present the different [K/O] models (solid lines) we fit against the transmission spectrum at R=300 (black and white points). We represent each model's respective fit in the orange shading. (https://github.com/afeinstein20/wasp39b_niriss_paper/blob/main/scripts/edfigure8.py)

**Extended Data Table 1: An outline of reduction and fitting pipelines used to produce transmission spectra for WASP-39b with NIRISS-SOSS**. Size of the box aperture is listed in parentheses when appropriate. All spectra will be made publicly available.

| Pipeline | JDox bkg model | F277W filter | 1/*f* removal | Spectral extraction | Fitting process | Limb darkening |
|---|---|---|---|---|---|---|
| nirHiss | x | x | int | Box (24) | chromat_fitting; MCMC | Quadratic, fit; priors from ExoTiC-LD |
| supreme-SPOON | x | | group | ATOCA | juliet; LM(t) | Quadratic, fit; priors from ExoTiC-LD |
| transitspectroscopy | x | | int | Box (30) | juliet; GP | Truncated normal; priors from limb-darkening |
| NAMELESS | x | | int | Box (30) | ExoTEP; MCMC | Quadratic; uniform priors |
| iraclis | | | int | Smoothed optimal | iraclis; MCMC/emcee | Quadratic; priors from ExoTETHyS |
| FIREFly | | | int | optimal | FIREFly; MCMC | Quadratic; uniform priors |

**Extended Data Table 2: White-light curve best-fit orbital parameters from Order 1.** Transit time ($t_0$) is presented with respect to $t_0$ - 2459787 [BJD].

| **Pipeline** | $t_o$ | $R_p/R_{star}$ | $a/R_{star}$ | $b$ |
|---|---|---|---|---|
| nirHiss | 0.556726 ± 0.000009 | 0.14602 ± 0.00011 | 11.372 ± 0.015 | 0.4536 ± 0.0029 |
| supreme-SPOON | 0.556742 ± 0.000015 | 0.14588 ± 0.00020 | $11.3795^{+0.0253}_{-0.0263}$ | $0.4530^{+0.0053}_{-0.0052}$ |
| transitspectroscopy | 0.556743 ± 0.000020 | 0.14530 ± 0.00035 | $11.388^{+0.028}_{-0.027}$ | $0.4496^{+0.0057}_{-0.0060}$ |
| NAMELESS | 0.556740 ± 0.000014 | 0.14587 ± 0.00018 | 11.400 ± 0.024 | $0.4510^{+0.0047}_{-0.0049}$ |
| iraclis | 0.556724 ± 0.000018 | $0.145083^{+0.000090}_{-0.000083}$ | 11.419 ± 0.026 | 0.4439 ± 0.0062 |
| FIREFly | 0.556737 ± 0.000020 | 0.145982 ± 0.00025 | 11.402 ± 0.033 | 0.4519 ± 0.0067 |

**Data Availability**

The raw data from this study are publicly available via the Space Science Telescope Institute's Mikulski Archive for Space Telescopes (https://archive.stsci.edu/). The data which was used to create all of the figures in this manuscript are freely available on Zenodo and GitHub (Zenodo Link; https://github.com/afeinstein20/wasp39b_niriss_paper). All additional data is available upon request.

**Code Availability**

The following are open-source pipelines written in Python that are available either through the Python Package Index (PyPI) or GitHub that were used throughout this work:
Eureka! (https://github.com/kevin218/Eureka);
nirHiss (https://github.com/afeinstein20/nirhiss);
supreme-SPOON (https://github.com/radicamc/supreme-spoon);
transitspectroscopy (https://github.com/nespinoza/transitspectroscopy/tree/dev);
iraclis (https://github.com/ucl-exoplanets/Iraclis);
juliet (https://github.com/nespinoza/juliet);
chromatic (https://github.com/zkbt/chromatic);

chromatic_fitting (https://github.com/catrionamurray/chromatic_fitting);
ExoTiC-LD[54, 121] (https://github.com/Exo-TiC/ExoTiC-LD);
ExoTETHyS[122] (https://github.com/ucl-exoplanets/ExoTETHyS);
PICASO[88,89] (https://github.com/natashabatalha/picaso);
Virga[94, 95] (https://github.com/natashabatalha/virga);
CHIMERA (https://github.com/mrline/CHIMERA);
PyMultiNest (https://github.com/JohannesBuchner/PyMultiNest);
MultiNest (https://github.com/JohannesBuchner/MultiNest)

**Methods References**

42. Bell, T. J. *et al.* Eureka!: An End-to-End Pipeline for JWST Time-Series Observations. *arXiv e-prints* arXiv:2207.03585 (2022).
43. van Dokkum, P. G. Cosmic-Ray Rejection by Laplacian Edge Detection. **113**, 1420–1427 (2001).
44. Craig, M. *et al.* Astropy/ccdproc: v1.3.0.post1. (2017) doi:10.5281/zenodo.1069648.
45. Darveau-Bernier, A. *et al.* ATOCA: an Algorithm to Treat Order Contamination. Application to the NIRISS SOSS Mode. **134**, 094502 (2022).
46. Radica, M. *et al.* APPLESOSS: A Producer of ProfiLEs for SOSS. Application to the NIRISS SOSS Mode. **134**, 104502 (2022).
47. Espinoza, N. *TransitSpectroscopy*. (Zenodo, 2022). doi:10.5281/zenodo.6960924.
48. Tsiaras, A. *et al.* A Population Study of Gaseous Exoplanets. **155**, 156 (2018).
49. Tsiaras, A. *et al.* A New Approach to Analyzing HST Spatial Scans: The Transmission Spectrum of HD 209458 b. **832**, 202 (2016).
50. Rustamkulov, Z., Sing, D. K., Liu, R. & Wang, A. Analysis of a JWST NIRSpec Lab Time Series: Characterizing Systematics, Recovering Exoplanet Transit Spectroscopy, and Constraining a Noise Floor. **928**, L7 (2022).
51. Salvatier, J., Wiecki, T. V. & Fonnesbeck, C. Probabilistic programming in python using PyMC3. *PeerJ Computer Science* **2**, e55 (2016).
52. Claret, A. A new non-linear limb-darkening law for LTE stellar atmosphere models. Calculations for -5.0 <= log[M/H] <= +1, 2000 K <= $T_{eff}$ <= 50000 K at several surface gravities. **363**, 1081–1190 (2000).
53. Sing, D. K. Stellar limb-darkening coefficients for CoRot and Kepler. **510**, A21 (2010).
54. 54. Laginja, I. & Wakeford, H. ExoTiC-ISM: A Python package for marginalised exoplanet transit parameters across a grid of systematic instrument models. *The Journal of Open Source Software* **5**, 2281 (2020).
55. Gelman, A. & Rubin, D. B. Inference from iterative simulation using multiple sequences. *Statistical Science* **7**, 457–472 (1992).
56. Vehtari, A., Gelman, A., Simpson, D., Carpenter, B. & Bürkner, P.-C. Rank-normalization, folding, and localization: An improved $\hat{R}$ for assessing convergence of MCMC. *arXiv e-prints* arXiv:1903.08008 (2019).

57. Espinoza, N., Kossakowski, D. & Brahm, R. Juliet: A versatile modelling tool for transiting and non-transiting exoplanetary systems. *Monthly Notices of the Royal Astronomical Society* **490**, 2262–2283 (2019).
58. Kipping, D. M. Efficient, uninformative sampling of limb darkening coefficients for two-parameter laws. *Monthly Notices of the Royal Astronomical Society* **435**, 2152–2160 (2013).
59. Maciejewski, G. *et al.* New Transit Observations for HAT-P-30 b, HAT-P-37 b, TrES-5 b, WASP-28 b, WASP-36 b and WASP-39 b. *J Acta Astronomica* **66**, 55–74 (2016).
60. Foreman-Mackey, D., Agol, E., Ambikasaran, S. & Angus, R. Fast and Scalable Gaussian Process Modeling with Applications to Astronomical Time Series. **154**, 220 (2017).
61. Espinoza, N. & Jordán, A. Limb darkening and exoplanets - II. Choosing the best law for optimal retrieval of transit parameters. **457**, 3573–3581 (2016).
62. Espinoza, N. & Jordán, A. Limb darkening and exoplanets: testing stellar model atmospheres and identifying biases in transit parameters. **450**, 1879–1899 (2015).
63. Howarth, I. D. On stellar limb darkening and exoplanetary transits. **418**, 1165–1175 (2011).
64. Benneke, B. *et al.* Water Vapor and Clouds on the Habitable-zone Sub-Neptune Exoplanet K2-18b. *The Astrophysical Journal Letters* **887**, L14 (2019).
65. Mandel, K. & Agol, E. Analytic Light Curves for Planetary Transit Searches. **580**, L171–L175 (2002).
66. Kreidberg, L. Batman : BAsic Transit Model cAlculatioN in Python. *Publications of the Astronomical Society of the Pacific* **127**, 1161–1165 (2015).
67. Foreman-Mackey, D., Hogg, D. W., Lang, D. & Goodman, J. Emcee: The MCMC hammer. *Publications of the Astronomical Society of the Pacific* **125**, 306 (2013).
68. Tsiaras, A. *et al.* pylightcurve: Exoplanet lightcurve model. ascl:1612.018 (2016).
69. Morello, G. *et al.* The ExoTETHyS Package: Tools for Exoplanetary Transits around Host Stars. **159**, 75 (2020).
70. Hellier, C. *et al.* Transiting hot Jupiters from WASP-South, Euler and TRAPPIST: WASP-95b to WASP-101b. **440**, 1982–1992 (2014).
71. Virtanen, P. *et al.* scipy/scipy: SciPy 1.5.3. (2020) doi:10.5281/zenodo.4100507.
72. Tremblin, P. *et al.* Fingering Convection and Cloudless Models for Cool Brown Dwarf Atmospheres. **804**, L17 (2015).
73. Drummond, B. *et al.* The effects of consistent chemical kinetics calculations on the pressure-temperature profiles and emission spectra of hot Jupiters. **594**, A69 (2016).
74. Goyal, J. M. *et al.* A library of ATMO forward model transmission spectra for hot Jupiter exoplanets. **474**, 5158–5185 (2018).
75. Goyal, J. M. *et al.* A library of self-consistent simulated exoplanet atmospheres. **498**, 4680–4704 (2020).
76. Barber, R. J., Tennyson, J., Harris, G. J. & Tolchenov, R. N. A high-accuracy computed water line list. **368**, 1087–1094 (2006).

77. Yurchenko, S. N. & Tennyson, J. ExoMol line lists - IV. The rotation-vibration spectrum of methane up to 1500 K. **440**, 1649–1661 (2014).
78. Tashkun, S. A. & Perevalov, V. I. CDSD-4000: High-resolution, high-temperature carbon dioxide spectroscopic databank. *Journal of Quantitative Spectroscopy and Radiative Transfer* **112**, 1403–1410 (2011).
79. Rothman, L. S. *et al.* HITEMP, the high-temperature molecular spectroscopic database. **111**, 2139–2150 (2010).
80. Ryabchikova, T. *et al.* A major upgrade of the VALD database. **90**, 054005 (2015).
81. Hauschildt, P. H., Allard, F. & Baron, E. The NextGen Model Atmosphere Grid for 3000<=$T_{eff}$ <=10,000 K. **512**, 377–385 (1999).
82. Barman, T. S., Hauschildt, P. H. & Allard, F. Irradiated Planets. **556**, 885–895 (2001).
83. Lothringer, J. D. & Barman, T. S. The PHOENIX Exoplanet Retrieval Algorithm and Using $H^{-}$ Opacity as a Probe in Ultrahot Jupiters. **159**, 289 (2020).
84. Rothman, L. S. *et al.* The HITRAN 2008 molecular spectroscopic database. *Journal of Quantitative Spectroscopy and Radiative Transfer* **110**, 533–572 (2009).
85. Kurucz, R. & Bell, B. Atomic Line Data. *Atomic Line Data (R.L. Kurucz and B. Bell) Kurucz CD-ROM No. 23. Cambridge* **23**, (1995).
86. McKay, C. P., Pollack, J. B. & Courtin, R. The thermal structure of Titan's atmosphere. **80**, 23–53 (1989).
87. Marley, M. S. & McKay, C. P. Thermal Structure of Uranus' Atmosphere. **138**, 268–286 (1999).
88. Batalha, N. E., Marley, M. S., Lewis, N. K. & Fortney, J. J. Exoplanet Reflected-light Spectroscopy with PICASO. **878**, 70 (2019).
89. Mukherjee, S., Batalha, N. E., Fortney, J. J. & Marley, M. S. PICASO 3.0: A One-Dimensional Climate Model for Giant Planets and Brown Dwarfs. *arXiv e-prints* arXiv:2208.07836 (2022).
90. Polyansky, O. L. *et al.* ExoMol molecular line lists XXX: a complete high-accuracy line list for water. **480**, 2597–2608 (2018).
91. Yurchenko, S. N., Amundsen, D. S., Tennyson, J. & Waldmann, I. P. A hybrid line list for $CH_4$ and hot methane continuum. **605**, A95 (2017).
92. Huang, X., Gamache, R., Freedman, R., Schwenke, D. & Lee, T. Reliable infrared line lists for 13 $CO_2$ isotopologues up to e′=18,000 $cm^{-1}$ and 1500 k, with line shape parameters. *Journal of Quantitative Spectroscopy & Radiative Transfer* **147**, 134–144 (2014).
93. Li, G. *et al.* Rovibrational Line Lists for Nine Isotopologues of the CO Molecule in the X $^1\Sigma^+$ Ground Electronic State. **216**, 15 (2015).
94. Ackerman, A. S. & Marley, M. S. Precipitating Condensation Clouds in Substellar Atmospheres. **556**, 872–884 (2001).
95. Rooney, C. M., Batalha, N. E., Gao, P. & Marley, M. S. A New Sedimentation Model for Greater Cloud Diversity in Giant Exoplanets and Brown Dwarfs. **925**, 33 (2022).

96. Mai, C. & Line, M. R. Exploring Exoplanet Cloud Assumptions in JWST Transmission Spectra. **883**, 144 (2019).
97. Bohren, C. F. & Huffman, D. R. *Absorption and scattering of light by small particles*. (1983).
98. Arcangeli, J. *et al.* $H^-$ Opacity and Water Dissociation in the Dayside Atmosphere of the Very Hot Gas Giant WASP-18b. **855**, L30 (2018).
99. Piskorz, D. *et al.* Ground- and Space-based Detection of the Thermal Emission Spectrum of the Transiting Hot Jupiter KELT-2Ab. **156**, 133 (2018).
100. Mansfield, M. *et al.* A unique hot Jupiter spectral sequence with evidence for compositional diversity. *Nature Astronomy* **5**, 1224–1232 (2021).
101. The JWST Transiting Exoplanet Community Early Release Science Team *et al.* Identification of carbon dioxide in an exoplanet atmosphere. *arXiv e-prints* arXiv:2208.11692 (2022).
102. Fortney, J. J. The effect of condensates on the characterization of transiting planet atmospheres with transmission spectroscopy. **364**, 649–653 (2005).
103. Line, M. R. *et al.* A Systematic Retrieval Analysis of Secondary Eclipse Spectra. I. A Comparison of Atmospheric Retrieval Techniques. **775**, 137 (2013).
104. Iyer, A. R. & Line, M. R. The Influence of Stellar Contamination on the Interpretation of Near-infrared Transmission Spectra of Sub-Neptune Worlds around M-dwarfs. **889**, 78 (2020).
105. Feroz, F., Hobson, M. P. & Bridges, M. MULTINEST: an efficient and robust Bayesian inference tool for cosmology and particle physics. **398**, 1601–1614 (2009).
106. Buchner, J. *et al.* X-ray spectral modelling of the AGN obscuring region in the CDFS: Bayesian model selection and catalogue. **564**, A125 (2014).
107. Richard, C. *et al.* New section of the HITRAN database: Collision-induced absorption (CIA). **113**, 1276–1285 (2012).
108. Freedman, R. S. *et al.* Gaseous Mean Opacities for Giant Planet and Ultracool Dwarf Atmospheres over a Range of Metallicities and Temperatures. **214**, 25 (2014).
109. Azzam, Ala'a. A. A., Tennyson, J., Yurchenko, S. N. & Naumenko, O. V. ExoMol molecular line lists - XVI. The rotation-vibration spectrum of hot $H_2S$. **460**, 4063–4074 (2016).
110. Barber, R. J. *et al.* ExoMol line lists - III. An improved hot rotation-vibration line list for HCN and HNC. **437**, 1828–1835 (2014).
111. Kramida, A., Yu. Ralchenko, Reader, J. & and NIST ASD Team. (2018).
112. Allard, N. F., Spiegelman, F., Leininger, T. & Molliere, P. New study of the line profiles of sodium perturbed by $H_2$. **628**, A120 (2019).
113. Allard, N. F., Spiegelman, F. & Kielkopf, J. F. K-$H_2$ line shapes for the spectra of cool brown dwarfs. **589**, A21 (2016).
114. Gharib-Nezhad, E. *et al.* EXOPLINES: Molecular Absorption Cross-section Database for Brown Dwarf and Giant Exoplanet Atmospheres. **254**, 34 (2021).

115. Grimm, S. L. *et al.* HELIOS-K 2.0 Opacity Calculator and Open-source Opacity Database for Exoplanetary Atmospheres. **253**, 30 (2021).
116. Lecavelier Des Etangs, A., Pont, F., Vidal-Madjar, A. & Sing, D. Rayleigh scattering in the transit spectrum of HD 189733b. **481**, L83–L86 (2008).
117. Ohno, K. & Kawashima, Y. Super-Rayleigh Slopes in Transmission Spectra of Exoplanets Generated by Photochemical Haze. **895**, L47 (2020).
118. Perez-Becker, D. & Showman, A. P. Atmospheric Heat Redistribution on Hot Jupiters. **776**, 134 (2013).
119. Komacek, T. D. & Showman, A. P. Atmospheric Circulation of Hot Jupiters: Dayside-Nightside Temperature Differences. **821**, 16 (2016).
120. Zhang, X. Atmospheric regimes and trends on exoplanets and brown dwarfs. *Research in Astronomy and Astrophysics* **20**, 099 (2020).
121. Wakeford, H. & Grant, D. *Exo-TiC/ExoTiC-LD: ExoTiC-LD v2.1 zenodo release*. (Zenodo, 2022). doi:10.5281/zenodo.6809899.
122. Morello, G. *et al.* ExoTETHyS: Tools for Exoplanetary Transits around host stars. *The Journal of Open Source Software* **5**, 1834 (2020).
123. Bradley, L. *et al.* *Astropy/photutils: 1.0.0.* (Zenodo, 2020). doi:10.5281/zenodo.4044744.

**Acknowledgements**

This work is based on observations made with the NASA/ESA/CSA James Webb Space Telescope. The data were obtained from the Mikulski Archive for Space Telescopes at the Space Telescope Science Institute, which is operated by the Association of Universities for Research in Astronomy, Inc., under NASA contract NAS 5-03127 for JWST. These observations are associated with program #1366. Support for this program was provided by NASA through a grant from the Space Telescope Science Institute. ADF acknowledges support by the National Science Foundation Graduate Research Fellowship Program under Grant No. DGE-1746045. MR acknowledges support from the National Sciences and Engineering Research Council of Canada, the Fonds de Recherche du Québec, as well as the Institut de Recherche sur les Exoplanétes. This research made use of `ccdproc`, an Astropy package for image reduction[44]. This research made use of Photutils, an Astropy package for the detection and photometry of astronomical sources[123].

**Author contributions**

All authors played a significant role in producing the results of this manuscript. Those roles include: developing the original JWST Early Release Science proposal, management of the project, definition of the target list, designing the observation plan, reduction and analysis of the data, theoretical modelling, and preparation of this manuscript.

Some specific contributions are listed as follows. NMB, JLB, and KBS provided overall program leadership and management. ADF, MR, LW and JLB led the efforts for this manuscript. DS, EK,

HRW, IC, JLB, KBS, LK, MLM, MRL, NMB, VP, and ZBT made significant contributions to the design of the program. KBS generated the observing plan with input from the team. NE provided instrument expertise. BB, EK, HRW, IC, JLB, LK, MLM, MRL, NMB, and ZBT led or co-led working groups and/or contributed to significant strategic planning efforts like the design and implementation of the pre-launch Data Challenges. AC, DS, ES, NE, NG, VP generated simulated data for pre-launch testing of methods. ADF, MR, LW, CAM, KO, JLB, and JT contributed significantly to the writing of this manuscript, along with contributions in the Methods from LPC, BB, ZR, AS, and AT. ADF, MR, CAM, LPC, NE, ZR, AS, and AT contributed to the development of data analysis pipelines and/or provided the data analysis products used in this analysis i.e., reduced the data, modelled the light curves, and/or produced the planetary spectrum. LW, KO, MRL, and SM generated theoretical model grids for comparison with data. ADF, LW, and KO generated figures for this manuscript. JB, LDS, JJF, PG, HAK, RJM, TME, BVR, and VP provided significant feedback to the manuscript coordinating comments from all other authors.

**Competing interests** The authors declare no competing interests.

**Author Affiliations**

[1] Department of Astronomy & Astrophysics, University of Chicago, Chicago, IL, USA

[2] NSF Graduate Research Fellow

[3] Department of Physics and Institute for Research on Exoplanets, Université de Montréal, Montreal, QC, Canada

[4] School of Earth and Space Exploration, Arizona State University, Tempe, AZ, USA

[5] NHFP Sagan Fellow

[6] Department of Astrophysical and Planetary Sciences, University of Colorado, Boulder, CO, USA

[7] Department of Astronomy & Astrophysics, University of California, Santa Cruz, Santa Cruz, CA, USA

[8] Space Telescope Science Institute, Baltimore, MD, USA

[9] Department of Physics & Astronomy, Johns Hopkins University, Baltimore, MD, USA

[10] Earth and Planets Laboratory, Carnegie Institution for Science, Washington, DC, USA

[11] Department of Earth and Planetary Sciences, Johns Hopkins University, Baltimore, MD, USA

[12] Department of Physics and Astronomy, University College London, United Kingdom

[13] INAF - Osservatorio Astrofisico di Arcetri, Largo E. Fermi 5, Firenze, Italy

[14] School of Physical Sciences, The Open University, Milton Keynes, UK

[15] Division of Geological and Planetary Sciences, California Institute of Technology, Pasadena, CA, USA

[16] Department of Astronomy, University of Michigan, Ann Arbor, MI, USA

[17] Department of Astronomy and Carl Sagan Institute, Cornell University, Ithaca, NY, USA

[18] Max Planck Institute for Astronomy, Heidelberg, Germany

[19] Department of Earth, Atmospheric and Planetary Sciences, Massachusetts Institute of Technology, Cambridge, MA, USA
[20] Kavli Institute for Astrophysics and Space Research, Massachusetts Institute of Technology, Cambridge, MA, USA
[21] 51 Pegasi b Fellow
[22] Atmospheric, Oceanic and Planetary Physics, Department of Physics, University of Oxford, Oxford, UK
[23] Université Côte d'Azur, Observatoire de la Côte d'Azur, CNRS, Laboratoire Lagrange, France
[24] Astrobiology Program, UC Santa Cruz, Santa Cruz, CA, USA
[25] European Space Agency, Space Telescope Science Institute, Baltimore, MD, USA
[26] School of Physics, Trinity College Dublin, Dublin, Ireland
[27] School of Earth and Planetary Sciences (SEPS), National Institute of Science Education and Research (NISER), HBNI, Odisha, India
[28] Center for Astrophysics | Harvard & Smithsonian, Cambridge, MA, USA
[29] Department of Physics, Utah Valley University, Orem, UT, USA
[30] Leiden Observatory, University of Leiden, Leiden, The Netherlands
[31] SRON Netherlands Institute for Space Research, Leiden, the Netherlands
[32] Universitäts-Sternwarte, Ludwig-Maximilians-Universität München, München, Germany
[33] Exzellenzcluster Origins, Garching, Germany
[34] Lunar and Planetary Laboratory, University of Arizona, Tucson, AZ, USA.
[35] Instituto de Astrofísica de Canarias (IAC), Tenerife, Spain
[36] Departamento de Astrofísica, Universidad de La Laguna (ULL), Tenerife, Spain
[37] INAF- Palermo Astronomical Observatory, Piazza del Parlamento, Palermo, Italy
[38] Johns Hopkins APL, Laurel, MD, USA
[39] School of Physics, University of Bristol, Bristol, UK
[40] Centre for Exoplanets and Habitability, University of Warwick, Coventry, UK
[41] Department of Physics, University of Warwick, Coventry, UK
[42] NASA Ames Research Center, Moffett Field, CA, USA
[43] BAER Institute, NASA Ames Research Center, Moffet Field, CA, USA
[44] Department of Physics, New York University Abu Dhabi, Abu Dhabi, UAE
[45] Center for Astro, Particle and Planetary Physics (CAP3), New York University Abu Dhabi, Abu Dhabi, UAE
[46] Department of Physics & Astronomy, University of Kansas, Lawrence, KS, USA
[47] Instituto de Astrof\'{\i}sica, Universidad Andres Bello, Santiago, Chile
[48] N\'ucleo Milenio de Formaci\'on Planetaria (NPF), Chile
[49] Centro de Astrof\'{\i}sica y Tecnolog\'{\i}as Afines (CATA), Casilla 36-D, Santiago, Chile
[50] School of Physics and Astronomy, University of Leicester, Leicester
[51] Centre for Exoplanet Science, University of St Andrews, St Andrews, UK
[52] INAF – Osservatorio Astrofisico di Torino, Pino Torinese, Italy
[53] Space Research Institute, Austrian Academy of Sciences, Graz, Austria
[54] Institute of Astronomy, Department of Physics and Astronomy, KU Leuven, Leuven, Belgium

[55] Anton Pannekoek Institute for Astronomy, University of Amsterdam, Amsterdam, The Netherlands
[56] Planetary Sciences Group, Department of Physics and Florida Space Institute, University of Central Florida, Orlando, Florida, USA
[57] ARTORG Center for Biomedical Engineering, University of Bern, Bern, Switzerland
[58] Institute for Astrophysics, University of Vienna, Vienna, Austria
[59] Department of Astronomy, University of Maryland, College Park, MD, USA
[60] Department of Physics, Imperial College London, London, UK
[61] Imperial College Research Fellow
[62] California Institute of Technology, IPAC, Pasadena, CA, USA
[63] Université Paris-Saclay, Université Paris Cité, CEA, CNRS, AIM, Gif-sur-Yvette, France
[64] Département d'Astronomie, Université de Genève, Sauverny, Switzerland
[65] Department of Physics, University of Rome "Tor Vergata", Rome, Italy
[66] INAF - Turin Astrophysical Observatory, Pino Torinese, Italy
[67] Steward Observatory, University of Arizona, Tucson, AZ, USA
[68] Department of Physics and Astronomy, Faculty of Environment Science and Economy, University of Exeter, EX4 4QL, UK.
[69] Astronomy Department and Van Vleck Observatory, Wesleyan University, Middletown, CT, USA
[70] Institute of Astronomy, University of Cambridge, Cambridge, UK
[71] Universidad Adolfo Ibáñez: Penalolen, Santiago, CL
[72] Maison de la Simulation, CEA, CNRS, Univ. Paris-Sud, UVSQ, Université Paris-Saclay, Gif-sur-Yvette, France
[73] Université de Paris Cité and Univ Paris Est Creteil, CNRS, LISA, Paris, France
[74] Astrophysics and Planetary Sciences Department, University of Colorado Boulder, Boulder, CO, USA
[75] Department of Earth and Planetary Sciences, University of California Santa Cruz, Santa Cruz, California, USA